\documentclass[epj]{svjour}
\usepackage{graphics}

\usepackage[dvipsnames]{xcolor}
\usepackage{graphicx,color}
\usepackage{enumitem}
\usepackage{braket}
\usepackage{slashed}
\usepackage[utf8]{inputenc}
\usepackage{hyperref}
\usepackage{float}
\usepackage{caption}
\usepackage{subcaption}
\usepackage{makecell}
\usepackage{epsfig}
\usepackage{amsmath,amssymb}
\usepackage[normalem]{ulem}
\usepackage{tikz}
\usetikzlibrary{shapes.geometric, arrows}
\hypersetup{
	colorlinks=true,
	citecolor=black,
	filecolor=black,
	linkcolor=blue,
	urlcolor=blue
}

\def\KK{{\scriptscriptstyle KK}}

\def\LO{{\scriptscriptstyle LO}}
\def\NLO{{\scriptscriptstyle NLO}}
\def\NNLO{{\scriptscriptstyle NNLO}}
\def\DBI{{\scriptscriptstyle DBI}}
\def\CS{{\scriptscriptstyle CS}}

\begin{document}

\title{Back to the origins of brane-antibrane inflation}

\author{Michele Cicoli\inst{1,2} \and Christopher Hughes\inst{3} \and Ahmed Rakin Kamal\inst{4,5} \and Francesco Marino\inst{1} \and Fernando Quevedo\inst{3,6,7} \and Mario Ramos-Hamud\inst{3} \and Gonzalo Villa\inst{3}
}                     

\institute{Dipartimento di Fisica e Astronomia, Universit\`a di Bologna, via Irnerio 46, 40126 Bologna, Italy
\and INFN, Sezione di Bologna, viale Berti Pichat 6/2, 40127 Bologna, Italy 
\and 
DAMTP,University of Cambridge, Wilberforce Road,  Cambridge, CB3 0WA, UK
\and
Department of Theoretical Physics and Astrophysics, Masaryk University, Kotlářská 2, CS-61137 Brno, Czechia
\and 
Department of Mathematics and Natural Sciences, BRAC University, 66 Mohakhali, Dhaka 1212, Bangladesh
\and
CERN Theory Group, CH-1211, Geneva, Switzerland
\and
New York University, Abu Dhabi, PO Box 129188 Saadiyat Island, Abu Dhabi, UAE
}

\mail{michele.cicoli@unibo.it}

\date{Received: date / Revised version: date}

\abstract{We study a new framework for brane-antibrane inflation where moduli stabilisation relies purely on perturbative corrections to the effective action. This guarantees that the model does not suffer from the eta-problem. The inflationary potential has two contributions: the tension of an antibrane at the tip of a warped throat, and its Coulomb interaction with a mobile brane. This represents the first realisation of the original idea of brane-antibrane inflation, as opposed to inflection point inflation which arises when the moduli are fixed with non-perturbative effects. Moreover, we formulate the brane-antibrane dynamics as an F-term potential of a nilpotent superfield in a manifestly supersymmetric effective theory. We impose compatibility with data and consistency conditions on control over the approximations and find that slow-roll inflation can occur in a large region of the underlying parameter space. The scalar spectral index is in agreement with data and the tensor-to-scalar ratio is beyond current observational reach. Interestingly, after the end of inflation the volume mode can, but does not need to, evolve towards a late-time minimum at larger values.
\PACS{
      {PACS-key}{discribing text of that key}   \and
      {PACS-key}{discribing text of that key}
     } 
}

\titlerunning{Back to the origins of brane-antibrane inflation}
\authorrunning{M. Cicoli et al}
\maketitle

\section{Introduction}

After three decades of precision cosmology, inflation stands as the standard realisation of early universe cosmology. Despite hundreds of models of potential inflationary scenarios, observations have restricted substantially the general structure of the corresponding inflationary potentials. Observationally favoured scenarios \cite{Planck:2018jri,Martin:2024qnn} correspond to concave potentials such as: 
\begin{equation}
V(\varphi)= C_0\left(1-\frac{C_1}{\varphi^n}\right)
\quad \text{or} \quad V(\varphi)=C_2\left(1-C_3\,e^{-m\varphi}\right),
\label{GenPot0}
\end{equation}
with $C_i$, $n$ and $m$ positive constants. A broad region of these parameters allows for a slow-roll dynamics of the scalar field which gives rise to realistic inflation. These potentials are clearly favoured over convex potentials such as $V(\varphi)\propto \varphi^p$. They also differ among themselves mainly in the prediction for the tensor-to-scalar ratio $r$ since exponential potentials tend to yield larger values of $r$ as compared to inverse power-law potentials.\footnote{For instance $r\simeq 8\varphi^2\eta^2/(n-1)^2$ for inverse power-law potentials, while $r\simeq 8\eta^2/c^2$ for exponential potentials, with $\eta$ the second slow-roll parameter (see for example \cite{Burgess:2014tja}).} In this respect, their difference may be confronted with observations in the not too far future.

It is then important to realise UV complete models of inflation with these properties. Over the years several different proposals for inflationary potentials derived from string theory have been put forward (see for instance \cite{Cicoli:2023opf} for a recent review). Most of them are disfavoured by observations because they predict either a too small scalar spectral index $n_s$ or a too big tensor-to-scalar ratio $r$. However there are also concrete string realisations of the inflationary models (\ref{GenPot0}) which fit the data rather well, such as Loop Blow-up Inflation \cite{Bansal:2024uzr} for the case of an inverse power-law potential with $n=2/3$, and Fibre Inflation \cite{Cicoli:2008gp,Broy:2015zba,Cicoli:2016chb,Cicoli:2016xae,Cicoli:2017axo,Cicoli:2020bao} for the exponential case with $m=1/\sqrt{3}$.\footnote{The scalar potential for Fibre Inflation is very similar to the one of the popular Starobinsky model. Their difference can in principle be tested due to their slightly different prediction for $r$. Note that Fibre Inflation has a string theory derivation whereas a UV complete version of the Starobinsky model has not been found. It has even been argued that it cannot be UV completed \cite{Burgess:2016owb,Brinkmann:2023eph,Lust:2023zql}.} 

Historically, following the seminal paper \cite{Dvali:1998pa}, the first example of an inverse power-law potential in string cosmology has been actually proposed in the original formulation of brane-antibrane inflation which features $n=4$ \cite{Burgess:2001fx,Dvali:2001fw}. In that case the term proportional to $C_0$ is generated by the tension of an $\overline{\rm D3}$-brane, while the term proportional to $C_1$ corresponds to the Coulomb attraction between an $\overline{\rm D3}$-brane and a mobile D3-brane. 

This scenario is very appealing since it not only had a concrete string theory candidate for inflation (the brane separation), it also had a very stringy way to end inflation by means of an open string mode that becomes tachyonic at a critical separation \cite{Burgess:2001fx} therefore realising a version of hybrid inflation. However, it was soon realised \cite{Burgess:2001fx,Dvali:2001fw} that to get viable slow-roll inflation the distance between the brane and the antibrane should be larger than the typical size of the extra-dimensions. This problem was overcome in \cite{Kachru:2003sx} where the authors placed the $\overline{\rm D3}$-brane at the tip of a highly warped throat. The authors of the same paper however pointed out that a stable inflationary trajectory requires to stabilise the modulus corresponding to the volume of the extra dimensions. In turn, the dynamics responsible to fix the volume mode can induce inflaton-dependent contributions to the scalar potential which in general ruin its flatness. 

This is the famous $\eta$-problem that plagues several inflationary models in supergravity and string theory. In particular, after including these new $\varphi$-dependent contributions, the resulting potential does not feature anymore an inverse power-law behaviour with $n=4$. A more refined analysis \cite{Baumann:2007ah} showed that concrete models of brane-antibrane inflation can have enough tuning freedom to realise inflection point inflation. This is however a scenario that is very different from the original one proposed in \cite{Burgess:2001fx,Dvali:2001fw} and, above all, it is disfavoured by observations. Nevertheless, when comparing general models with observations, it has remained a practice in the literature to consider inverse power-law potentials still under the name of brane-antibrane inflation \cite{Planck:2018jri,Martin:2024qnn} even in the absence of a string realisation of these models once moduli stabilisation is taken into account. 

One of the purposes of this article is to remedy this situation by providing the first realisations of the original potential of brane-antibrane inflation in low-energy string models that include controlled moduli stabilisation. The key ingredient to avoid the $\eta$-problem is to fix the volume mode relying just on perturbative corrections to the effective action, following the recent proposal of \cite{Burgess:2022nbx} where the authors fixed the moduli with loop corrections in an RG-inspired setup from a bottom-up perspective. In our paper we expand on these results taking instead a more top-down perspective and stabilising the volume modulus using known $\alpha'$ and $g_s$ perturbative corrections to the K\"ahler potential of type IIB string compactifications. 

We first show that the leading perturbative corrections which break the no-scale structure of the low-energy effective field theory (EFT) have enough structure to generate a post-inflationary dS minimum for the volume mode at large volume and weak string coupling. This minimum is obtained by balancing different effects: $\mathcal{O}(g_s \alpha'^2)$ logarithmic redefinitions of the moduli \cite{Weissenbacher:2019mef,Weissenbacher:2020cyf,Klaewer:2020lfg}, $\mathcal{O}(\alpha'^3)$ corrections \cite{Becker:2002nn,Bonetti:2016dqh} and logarithmically enhanced $\mathcal{O}(g_s^2 \alpha'^3)$ contributions \cite{Antoniadis:2018hqy,Antoniadis:2019rkh}. We also check that additional perturbative corrections are subdominant and do not modify the leading order picture qualitatively. Let us stress that our stabilisation mechanism has mainly an illustrative purpose since we realised that dS vacua arise at perturbative level even more generically also for cases where some of the leading order effects that we considered are vanishing or absent by construction.  

With this late-time volume stabilisation at hand, we then turn to consider brane-antibrane inflation where we manage to realise it for the first time with a fixed volume mode and within a manifestly supersymmetric EFT. This last requirement is achieved by expressing the brane-antibrane dynamics in terms of the F-term potential of a nilpotent superfield $X$. We also present two possible ways to stabilise the volume mode in a dS vacuum during inflation depending on the actual dependence of the K\"ahler potential $K$ on the nilpotent superfield. When $e^{-K/3}$ features a linear dependence on $X$, the leading order potential becomes a perfect square which admits a Minkowski vacuum that is uplifted to positive energies by the inclusion of subdominant $\mathcal{O}(\alpha'^3)$ corrections. In this case the early and late-time minima are located at different regions in field space, and so the volume mode is expected to evolve dynamically after the end of inflation, as proposed in \cite{Conlon:2022pnx,Apers:2022cyl,Apers:2024ffe}. On the other hand, when $e^{-K/3}$ does not have any linear dependence on $X$, as it is known to be the case at tree-level, the minima for the volume mode during and after inflation lie around the same region in moduli space with the difference that during inflation the tension of the antibrane uplifts the vacuum energy to larger values.

Finally we compare our results with observations and check that the EFT is under control by requiring to have the D3-brane at an appropriate distance from the tip of the throat, a gravitino mass below the warped string scale, a large enough volume to trust the validity of the dilute flux approximation, curvature corrections under control and no destabilisation of the conifold modulus. This allows us to determine the allowed regions of the underlying parameter space, showing that the original model of brane-antibrane inflation can be successfully realised without any $\eta$-problem and with enough e-foldings to solve the horizon and flatness problems while at the same time producing density perturbations in the observationally allowed region. 

This paper is organised as follows. In Sec. \ref{Sec2} we provide a brief historical overview of brane-antibrane inflation with technical details discussed in App. \ref{sec:coulomb-potential}. Sec. \ref{sec:late-time} describes instead the late-time stabilisation of the volume mode using leading perturbative contributions to $K$, while App. \ref{sec:Subleading Appendix Late Time} analyses the effect of subdominant corrections. Sec. \ref{sec:inflation-alpha} focuses on the realisation of brane-antibrane inflation and moduli stabilisation at early-times, devoting App. \ref{AppC} to the discuss of subleading terms. In Sec. \ref{sec:Parameter-Space-Everything} we plot the allowed regions of the UV parameter space where EFT and observational constraints are satisfied. Finally, the proposal of \cite{Burgess:2022nbx} is reviewed in detail in App. \ref{sec:RG Parameter Space}.

\section{A historical perspective on brane-antibrane inflation}
\label{Sec2}

The Coulomb interaction among a brane and an antibrane is known to vanish at large distances. This simple observation implies that the open string mode controlling the distance between the brane and the antibrane is a natural inflaton candidate since its potential becomes very shallow at large field values \cite{Burgess:2001fx,Dvali:2001fw}.\footnote{See \cite{Quevedo:2002xw} for an old but detailed review on the subject. Recent takes are available in \cite{Baumann:2014nda,Tye:2023rwa}.} In fact, the potential for a pair of D3/$\overline{\rm D3}$-branes separated by a distance $r$ takes the form:
\begin{equation}
V(r) = C_0\left(1-\frac{D_0}{r^4}\right)\,,  
\label{V(r)}
\end{equation}
where in terms of the D3-brane tension $T_3$, the string coupling $g_s$ and the string length $\ell_s = 2\pi\sqrt{\alpha'} = M_s^{-1}$:
\begin{equation}
C_0 \equiv 2 T_3  =  \frac{4\pi}{g_s} M_s^4\qquad\text{and}\qquad D_0 \equiv \frac{1}{2\pi^2 T_3}\,.
\end{equation}
Note that the string scale $M_s$ can be expressed in terms of the 4D Planck scale $M_p$ as:
\begin{equation}
M_s = \frac{g_s M_p}{\sqrt{4\pi\mathcal{V}_s}}\,,
\label{StringScaleOrig}
\end{equation}
where $\mathcal{V}_s$ denotes the Calabi-Yau volume in string frame measured in units of the string length $\ell_s$. The potential (\ref{V(r)}) however would become flat enough to drive inflation only at distances larger than the size of the extra dimensions \cite{Burgess:2001fx}. This can be easily seen by computing the second slow-roll parameter $\eta$:
\begin{equation}
\eta = M_p^2\,\frac{V_{\varphi\varphi}}{V}\simeq -\frac{10}{\pi^3}\frac{\mathcal{V}_s}{(r M_s)^6}\,,
\label{etaOriginal}
\end{equation}
where the derivatives are taken with respect to the canonically normalised inflaton $\varphi = \sqrt{T_3}\,r$. If the Calabi-Yau volume is isotropic, the maximum value that $r$ can take is $r_{\rm max} \simeq \mathcal{V}_s^{1/6} M_s^{-1}$, which implies:
\begin{equation}
|\eta|\gtrsim \frac{10}{\pi^3}\simeq 0.3\,,
\end{equation}
that is not small enough to sustain enough e-foldings of inflation. Potential way-outs involve anisotropic compactifications, as the one obtained in \cite{Cicoli:2011yy} after stabilising the K\"ahler moduli, or warped geometries. This last option has been considered in \cite{Kachru:2003sx} that focused on Calabi-Yau geometries with a warped throat described by a cone over a 5D base that is an $S^2$ fibration over an $S^3$. The $\overline{\rm D3}$-brane sits at the tip of the warped throat where its energy is minimised. Inflation is then driven by the open string mode which controls the radial distance of a D3-brane from the $\overline{\rm D3}$-brane at the conifold singularity. The resulting inflationary potential takes the same form as (\ref{V(r)}) but now with:
\begin{equation}
C_0 \equiv 2 T_3\,\mathcal{V}^{2/3}\,e^{-8\pi K/(3g_sM)}\,,
\end{equation}
and:
\begin{equation}
D_0 \equiv \frac{27}{32\pi^2 T_3}\,\mathcal{V}^{2/3}\,e^{-8\pi K/(3g_sM)}\,,
\end{equation}
where $M$ is the quantised value of the $F_3$ flux on the $S^3$ at the tip, while $K$ is the quantised value of the $H_3$ flux on the Poincaré dual 3-cycle. Moreover $\mathcal{V}$ is the Calabi-Yau volume in Einstein frame which is related to the same quantity in string frame as $\mathcal{V}_s=g_s^{3/2}\mathcal{V}$. The new $\eta$-parameter can now be easily very small due to the warping suppression factor:
\begin{equation}
|\eta|\simeq \frac{135}{8\pi^3}\frac{\mathcal{V}_s}{(r M_s)^6}\,\mathcal{V}^{2/3}\,e^{-8\pi K/(3g_sM)} \ll 1\,.
\end{equation}
This analysis however ignores the fact that inflation takes place in a compactified setup with closed string moduli that need to be stabilised during inflation. A crucial field is in particular the overall volume mode $\mathcal{V}$. The string scale $M_s$ can be expressed in terms of this modulus as:
\begin{equation}
M_s = \frac{g_s^{1/4} M_p}{\sqrt{4\pi\mathcal{V}}}\,.
\label{StringScale}
\end{equation}
This implies that the inflationary energy density $C_0$ depends on $\mathcal{V}$ as:
\begin{equation}
C_0 = \frac{M_p^4}{4\pi\mathcal{V}^{4/3}}\,e^{-8\pi K/(3g_sM)}\equiv \frac{\mathcal{C}_0}{\mathcal{V}^{4/3}}\,,
\label{Avol}
\end{equation}
and so the inflationary potential is a function of both $r$ and $\mathcal{V}$:
\begin{equation}
V_{\rm inf} (r,\mathcal{V}) =  \frac{\mathcal{C}_0}{\mathcal{V}^{4/3}}\left[1-\frac{\mathcal{D}_0}{\left(r M_\KK\right)^4}\right]\,,
\label{VVinf}
\end{equation}
where:
\begin{equation}
\mathcal{D}_0 \equiv \left(\frac{3}{4\pi}\right)^3\,e^{-8\pi K/(3g_sM)}\,,
\end{equation}
and $M_\KK$ is the Kaluza-Klein (KK) scale given in terms of the stabilised volume $\langle\mathcal{V}\rangle$ when the D3-brane is near the tip of the throat \cite{Baumann:2007ah}:
\begin{equation}
M_\KK = \frac{M_s}{\langle\mathcal{V}_s\rangle^{1/6}} = \frac{M_p}{\sqrt{4\pi}\,\langle\mathcal{V}\rangle^{2/3}}\,.   
\end{equation}
Written in terms of the canonically normalised inflaton $\varphi$, the inflationary potential (\ref{VVinf}) takes the same form as the power-law potential in (\ref{GenPot0}) since it becomes:
\begin{equation}
V_{\rm inf} (\varphi,\mathcal{V}) =  \frac{\mathcal{C}_0}{\mathcal{V}^{4/3}}\left(1-\frac{C_1}{\varphi^4}\right)\qquad\text{with}\qquad C_1\equiv \frac{\mathcal{D}_0 T_3^2}{M_\KK^4}\,.
\label{InflPot}
\end{equation}
Therefore the volume mode $\mathcal{V}$ needs to be stabilised during inflation in order to avoid a dangerous runaway in a direction orthogonal to the inflationary one. Ref. \cite{Kachru:2003sx} exploited non-perturbative corrections to the superpotential to fix $\mathcal{V}$. However these non-perturbative effects induce a large contribution to the inflaton mass, destroying the flatness of its potential. This is a manifestation of the infamous $\eta$-problem that plagues several attempts to realise inflation in supergravity and string theory. 

The reason is the fact that the holomorphic superfield $T$, which appears in the superpotential $W$, is different from the physical Calabi-Yau volume $\mathcal{V}$. In fact, the correct definition of the chiral coordinate $T$ is such that:
\begin{equation}
T+\bar{T} = \mathcal{V}^{2/3} + \gamma r^2\,,
\label{ChiralCoord}
\end{equation}
with $\gamma \simeq T_3\, \langle(T+\bar{T})\rangle$. Hence $T$-dependent non-perturbative corrections to $W$ generate a potential $V_{\rm np}$ that depends on $\mathcal{V}$ and $T+\bar{T}$, and so on both $\mathcal{V}$ and $r$, after using (\ref{ChiralCoord}). In fact, after fixing $(T-\bar{T})$, $V_{\rm np}$ looks like:
\begin{equation}
V_{\rm np}(r,\mathcal{V})=  \frac{1}{\mathcal{V}^{4/3}}\,U_{\rm np}(T+\bar{T})\,.
\end{equation}
Using (\ref{ChiralCoord}) and writing (\ref{VVinf}) as $V_{\rm inf}= \mathcal{V}^{-4/3}\,U_{\rm inf}(r)$, the total scalar potential hence becomes:
\begin{equation}
V_{\rm tot} = V_{\rm np} + V_{\rm inf} = \frac{1}{\left(T+\bar{T} - \gamma r^2  \right)^2}\left[U_{\rm np}(T+\bar{T})+U_{\rm inf}(r)\right]
\label{Vtot}
\end{equation}
Expanding for $\gamma r^2\ll (T+\bar{T})$, we find:
\begin{equation}
V_{\rm tot} = \frac{1}{\left(T+\bar{T} \right)^2}\left[U_{\rm np}(T+\bar{T})+U_{\rm inf}(r)\right] \left(1 +\frac{2 \gamma r^2}{(T+\bar{T})}  \right).
\end{equation}
At the end of inflation, the D3-brane annihilates with the $\overline{\rm D3}$-brane and $U_{\rm inf}\to 0$. Hence one is leftover only with the non-perturbative potential which admits an AdS vacuum. To match observations, one has to add an appropriate uplifting source $C_{\rm up}$, so that $U_{\rm np}(T+\bar{T})\to U_{\rm np}(T+\bar{T}) + C_{\rm up}$. This late-time potential has now a minimum at $\langle (T+\bar{T})\rangle$ where $\langle U_{\rm np} \rangle + C_{\rm up}\simeq 0$. During inflation, the shift of the K\"ahler modulus from its late-time minimum is negligible, and so the inflationary potential takes the form:
\begin{equation}
V_{\rm inf} = \frac{U_{\rm inf}(r)}{\langle(T+\bar{T}) \rangle^2} \left(1 +\frac{2 \gamma r^2}{\langle(T+\bar{T})\rangle}  \right).
\end{equation}
Trading $r$ for the canonically normalised inflaton $\varphi$, this potential reduces to:
\begin{equation}
V_{\rm inf} = V_0(\varphi)\left(1+\frac13\,\frac{\varphi^2}{M_p^2}\right)\,,
\end{equation}
where $V_0(\varphi)$ should be identified with the leading inflationary potential given by (\ref{InflPot}) with fixed $\mathcal{V}$. It is then straightforward to realise that the Planck-suppressed 6D operator $V_0(\varphi)\varphi^2$ induces a large contribution to the slow-roll parameter $\eta$ of order $\Delta\eta=2/3$ which ruins inflation. Ref. \cite{Kachru:2003sx} argued however that the prefactor of non-perturbative corrections to $W$ is in general a function of $\varphi$, and so one should have more precisely that $U_{\rm np}=U_{\rm np}(T+\bar{T},\varphi) = U_{\rm np}(T+\bar{T}) + U_{\rm sub}(T+\bar{T},\varphi)$ where we included the $\varphi$-dependence in a subleading term. Thus the expansion of the scalar potential should contain an additional term of the form:
\begin{equation}
V_{\rm inf} \simeq V_0(\varphi) \left( 1+\frac13\,\frac{\varphi^2}{M_p^2}+P(\varphi)\right)
\end{equation}
where:
\begin{equation}
P(\varphi)\equiv \frac{U_{\rm sub}(\langle(T+\bar{T})\rangle,\varphi)}{U_{\rm inf}(\varphi)}\,.
\label{NewPot}
\end{equation}
This potential is now completely different from the original one (\ref{InflPot}) which corresponds to $V_0(\varphi)$. The new potential (\ref{NewPot}) can in principle sustain inflection point inflation around $\varphi=\varphi_0$ if the new term proportional to $P(\varphi)$ leads to a correction to $\eta$ of order $\Delta \eta(\varphi_0)\simeq -2/3$ which would cancel the contribution from the dangerous Planck-suppressed 6D operator. Inflation can then take place only around this point, in a tuned scenario that turns out to be inflection point inflation \cite{Baumann:2007ah}. 

Let us also mention that in \cite{Burgess:2001fx,Dvali:2001fw,Kachru:2003sx,Baumann:2007ah} the brane-antibrane potential has been introduced by hand without a manifestly supersymmetric framework. Recent progress has however shown that the positive contribution from the $\overline{\rm D3}$-brane at the tip of the warped throat (the $C_0$ term shown in (\ref{Avol})) can arise as the F-term contribution of a nilpotent superfield $X$ \cite{Aparicio:2015psl}. Ref. \cite{Burgess:2022nbx} has argued that the same F-term contribution should also generate the Coulomb interaction (the term proportional to $\mathcal{D}_0$ in (\ref{VVinf})) once the superpotential contains a coupling between the nilpotent superfield $X$ and the D3-brane radial coordinate $r$. Ref. \cite{Burgess:2022nbx} has also sketched a moduli stabilisation scheme based purely on perturbative corrections inferred from general RG-running considerations from a bottom-up perspective. This approach would avoid the $\eta$-problem, as also originally pointed out in \cite{Kachru:2003sx}, since perturbative corrections to the K\"ahler potential depend just on $\mathcal{V}$, but not on the inflaton $r$. In what follows we shall analyse this idea in detail from a more top-down point of view, showing that known perturbative corrections to the K\"ahler potential of string compactifications can allow to fix the volume mode perturbatively.

\section{Late-time moduli stabilisation}
\label{sec:late-time}

We start our discussion at late-times, when the effects of the antibrane are not present due to brane-antibrane annihilation. The effective field theory contains only the volume modulus, and in this section we propose a new perturbative moduli stabilisation scenario using known $\alpha'$ and loop corrections to the low-energy action of type IIB string compactifications. The tree-level K\"ahler potential reads:
\begin{equation}
K_{\rm tree}= -3\ln \tau\,,
\label{Ktree}
\end{equation}
where $\tau$ denotes the real part of the K\"ahler modulus $T$, $\tau\equiv (T+\bar{T})$, which is the appropriate chiral coordinate of the $N=1$ EFT. Note that at tree-level $\tau$ is given in terms of the physical Calabi-Yau volume $\mathcal{V}$ as $\tau=\mathcal{V}^{2/3}$. Perturbative corrections\footnote{For perturbative corrections in type IIB in 10 dimensions, see \cite{Liu:2022bfg,Wulff:2021fhr,Wulff:2024mgu}.} to (\ref{Ktree}) arise at different orders in $\alpha'$ and $g_s$, as analysed systematically in \cite{Cicoli:2021rub,Burgess:2020qsc}. Here we just briefly summarise the behaviour of known corrections:
\begin{enumerate}
\item $\mathcal{O}(\alpha')$: No correction is known to arise at this perturbative order.

\item $\mathcal{O}(\alpha'^2)$: Ref. \cite{Grimm:2013gma} found that $N=1$ $\mathcal{O}(g_s\alpha'^2)$ effects do not correct $K$ but induce just moduli redefinitions of the form:
\begin{equation}
\tau=\mathcal{V}^{2/3} \quad \to \quad \tau=\mathcal{V}^{2/3} + c_0\,,
\end{equation}
where $c_0$ is a topological constant. On the other hand, \cite{Weissenbacher:2019mef,Weissenbacher:2020cyf} claimed that $\mathcal{O}(g_s\alpha'^2)$ effects should induce logarithmic redefinitions of the moduli which would correct $K$. A similar conclusion has been recently reached in \cite{Klaewer:2020lfg} which obtained:
\begin{equation}
\tau \quad\to\quad \tau -\alpha\ln\tau 
\end{equation}
that implies:
\begin{equation}
K=-3\ln\tau \quad\to\quad  K=-3\ln\left(\tau-\alpha\ln\tau\right),
\label{ModRedef}
\end{equation}
where $\alpha$ depends on the 1-loop $\beta$-function coefficient $\beta_0$ of the field theory living on D7-branes as $\alpha=\beta_0/(8\pi)$. Note that gauge threshold corrections on D7-branes have been studied also in \cite{Conlon:2009xf,Conlon:2009qa,Conlon:2009kt} which found however just a redefinition of blow-up moduli. Moreover, \cite{Berg:2005ja} computed actual $N=2$ $\mathcal{O}(g_s^2\alpha'^2)$ corrections to the K\"ahler potential in toroidal orientifolds which have been generalised to arbitrary Calabi-Yau backgrounds in \cite{Berg:2007wt}. They take the form:
\begin{equation}
K_{\mathcal{O}(g_s^2\alpha'^2)} \simeq \frac{g_s c_1}{\tau}\,,
\label{Kloop}
\end{equation}
where $c_1$ is a function of the complex structure moduli. Interestingly, \cite{Cicoli:2007xp} found that the contribution to the scalar potential of this correction, as for any $\mathcal{O}(\alpha'^2)$ term at any order in the $g_s$ expansion, vanishes since it enjoys an \emph{extended no-scale structure}. This is not surprising since (\ref{Kloop}) represents the first term of the expansion of a K\"ahler potential of the form:
\begin{equation}
K =-3\ln\left(\tau+k\right)\qquad\text{with}\qquad k\equiv -\frac{g_s c_1}{3}\,,    
\label{Kexact}
\end{equation}
that would respect an exact no-scale structure at all orders \cite{Burgess:2020qsc}. It is unknown at the moment if higher order loop corrections respect the expansion of (\ref{Kexact}) (in particular those of $\mathcal{O}(g_s^4\alpha'^4)$ that would be the next order). 

\item $\mathcal{O}(\alpha'^3)$: The K\"ahler potential receives several corrections at $\mathcal{O}(\alpha'^3)$ which look like:
\begin{eqnarray}
\label{Kalpha'3}
K_{\mathcal{O}(g_s^n\alpha'^3)} &=&
-2 \ln\left\{1 + \frac{\xi}{2 (g_s\tau)^{3/2}}\right. \\
&& \left.\left[1+ g_s^2 \left(c_2\left(1-\frac{3 T_7}{2}\ln\tau\right)+c_3\right)\right]\right\},
\nonumber
\end{eqnarray}
with $T_7=2\pi$ the D7-brane tension and:
\begin{equation}
\xi =-\frac{\zeta(3)}{2(2 \pi)^3 } \left( \chi(\rm{CY})+2\int_{\rm CY} D_{\rm O7}^3\right)\,\,\text{and}\,\,c_2 =\frac{2\zeta(2)}{\zeta(3)},
\label{xiDef}
\end{equation}
where $\chi({\rm CY})$ is the Calabi-Yau Euler number and $D_{\rm O7}$ the $(1,1)$-form dual to the divisor wrapped by the O7-plane. The first correction proportional to $\chi({\rm CY})$ is an $N=2$ $\mathcal{O}(\alpha'^3)$ effect \cite{Becker:2002nn,Bonetti:2016dqh}, while $N=1$ contributions at the same order induce an O7-dependent shift \cite{Minasian:2015bxa}. The term proportional to $c_2$ is an $N=2$ $\mathcal{O}(g_s^2\alpha'^3)$ correction \cite{Antoniadis:1997eg,Leontaris:2022rzj} with the logarithmic contribution arising only in the presence of a localised high curvature in the extra dimensions \cite{Antoniadis:2018hqy,Antoniadis:2019rkh}. On the other hand, the tiny constant $c_3\sim 10^{-4}$ controls $N=1$ effects at the same order \cite{Berg:2014ama,Haack:2015pbv}. There are also $N=2$ $\mathcal{O}(\alpha'^3)$ corrections which are non-K\"ahler and give directly a correction to the scalar potential that scales as an $($F-term$)^4$ \cite{Ciupke:2015msa,Grimm:2017okk}:
\begin{equation}
V_{F^4} = c_3 \sqrt{g_s}\frac{W_0^4}{\tau^{11/2}}\,,
\end{equation}
where $c_3$ is a small topological quantity that has been estimated to be of order $10^{-3}$-$10^{-4}$ \cite{Cicoli:2023njy}.\footnote{See also \cite{deAlwis:2012vp,Cicoli:2013swa} for studies on the  consistency of the superspace derivative
expansion.}

\item $\mathcal{O}(\alpha'^4)$: At this order of approximation, the relevant corrections arise at $N=2$ $\mathcal{O}(g_s^2\alpha'^4)$ order and read \cite{Berg:2005ja,Berg:2007wt}:
\begin{equation}
K_{\mathcal{O}(g_s^2\alpha'^4)} \simeq \frac{c_4}{\tau^2}\,,
\label{Kloop2}
\end{equation}
where $c_4$ is a function of the complex structure moduli. Ref. \cite{Gao:2022uop} clarified the origin of these effects as loops of Kaluza-Klein modes of open strings stretched between intersecting branes. Similar corrections should arise also from loops of closed strings \cite{Gao:2022uop,vonGersdorff:2005bf}.
\end{enumerate}

After this brief review of known perturbative corrections to the low-energy EFT of type IIB string compactification, we are now ready to study perturbative moduli stabilisation. Working at large volume, $\tau\gg 1$, and weak string coupling, $g_s\ll 1$, the leading effects that are sufficient to consider to find a Minkowski vacuum are the $\alpha$-dependent moduli redefinition at $\mathcal{O}(g_s\alpha'^2)$ shown in (\ref{ModRedef}), the $\mathcal{O}(\alpha'^3)$ correction and the logarithmically enhanced $\mathcal{O}(g_s^2\alpha'^3)$ term in (\ref{Kalpha'3}). The leading order contribution of these three corrections can be captured by a K\"ahler potential of the form:
\begin{equation}
K\simeq -3 \ln \left[\tau-\alpha\ln \tau +\frac{\xi}{3 c g_s^{3/2}\sqrt{\tau}} \left(c   -  g_s^2 \ln\tau  \right) \right],
\label{eq:kahler-pot}
\end{equation}
where:\footnote{According to \cite{Antoniadis:2018hqy}, $T_7$ should be considered as an effective D7-brane tension that could be tuned by varying the complex structure moduli. In our analysis we did not use this tuning freedom, which would however just make our solution more robust.} 
\begin{equation}
c\equiv \frac{\zeta(3)}{3\zeta(2) T_7} = \frac{\zeta(3)}{\pi^3}\simeq 0.04 \,.
\label{eq:xi-D-sigma-relation}
\end{equation}
As usual in type IIB flux compactifications, we consider the dilaton and the complex structure moduli fixed at tree-level by 3-form fluxes which induce a constant superpotential $W_0$. At this level of approximation, the K\"ahler modulus $T$ is flat due to the no-scale cancellation which is however broken by all corrections in (\ref{eq:kahler-pot}). Plugging this K\"ahler potential together with $W=W_0$ (which we consider to be real without loss of generality) into the general form of the $N=1$ supergravity F-term scalar potential, we find at leading order:
\begin{eqnarray}
\frac{V}{3 W_0^2}&=& e^K \left(\frac13 K^{T\overline{T}} K_T K_{\overline{T}} -1 \right) \nonumber \\
&\simeq& \frac{\alpha}{\tau^4} - \frac{\xi \sqrt{g_s}}{4 c \tau^{9/2}}\left(\ln\tau  -\frac{c}{g_s^2}\right).
\label{eq:late-time-pot}
\end{eqnarray}
For $\alpha=0$ and $\xi>0$, this potential clearly admits an AdS vacuum at $\tau \sim e^{c/g_s^2}\gg 1$ for $g_s\ll 1$, as already noticed in \cite{Antoniadis:2018hqy,Antoniadis:2019rkh}.\footnote{Perturbative AdS vacua have been also originally found in \cite{Berg:2005yu} by balancing $\mathcal{O}(\alpha'^3)$ effects against $\mathcal{O}(g_s^2\alpha'^4)$ string loops with tuned coefficients.} Note that if there is no logarithmic correction, i.e. $c\to \infty$, due to the absence of a high curvature region localised in the extra dimensions, a similar logarithmic contribution (scaling as $(\ln\tau)^{3/2}/\sqrt{g_s}$) could instead arise, as in standard LVS models, from integrating out a blow-up mode which supports non-perturbative corrections \cite{Balasubramanian:2005zx}. On the other hand, if $\alpha$ is positive and sufficiently small, as expected from a loop-suppressed factor, the $\alpha$-dependent term in (\ref{eq:late-time-pot}) has the right volume scaling to behave as an uplifting contribution which can lead to a Minkowski or dS vacuum. Let us stress that, even if $\alpha=0$, the uplifting could come from several different sources like, for example, an $\overline{\rm D3}$-brane in a different throat from the inflationary one, T-branes \cite{Cicoli:2015ylx}, non-zero F-terms of the complex structure moduli \cite{Gallego:2017dvd}, D-terms \cite{Burgess:2003ic} or dilaton-dependent non-perturbative effects at singularities \cite{Cicoli:2012fh}.

\begin{figure*}[t]
\centering
\includegraphics[scale=1]{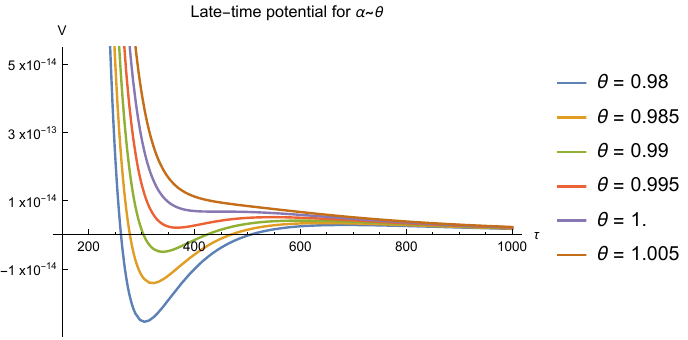}
\caption{Late-time potential for different values of the uplifting parameter $\theta$ setting $\xi=g_s =0.1$ and $T_7=2\pi$. By increasing $\theta$ we move from AdS to dS and finally to a runaway. In particular the red curve shows a dS minimum at $\tau_{\rm min} \sim 370$ and $V_{\rm min} \ll 10^{-14}$ for an appropriate choice of $\theta$.}
\label{fig:late-time-mins-plot-maintext}
\end{figure*}

In what follows we shall therefore focus on $\alpha>0$. In this case the potential \eqref{eq:late-time-pot} behaves asymptotically as follows: $V\to + \infty$ for $\tau\to 0$, and $V\to 0^+$ for $\tau\to\infty$. Hence it is either a runaway (for large values of $\alpha$) or it features a minimum and a maximum at larger field values (for smaller values of $\alpha$). We shall therefore focus on the second case where the minimum and the maximum are located at:
\begin{equation}
\tau_{\rm min}   = \lambda_0\, e^{\frac{c}{g_s^2}}\qquad\text{and}\qquad \tau_{\rm max}   = \lambda_{-1}\, e^{\frac{c}{g_s^2}}\,,
\label{eq:late-time-extrema-alpha-subbed}
\end{equation}
with:
\begin{equation}
\lambda_k \equiv e^{\frac29}\,e^{- 2 \mathcal{W}_k\left(-\theta/e\right)}\,,
\end{equation}
where:
\begin{equation}
\theta \equiv \left(\frac{16 c\,e^{\frac{10}{9}}}{9 \xi \sqrt{g_s}}\right) \alpha \, e^{\frac{c}{2 g_s^2}}\,,
\label{ThetaDef}
\end{equation}
and $\mathcal{W}_k(x)$ with $k=0,-1$  are respectively the $0$- and $(-1)$-branches of the Lambert function $\mathcal{W}_k(x)$ defined as  $\mathcal{W}_k(x) e^{\mathcal{W}_k(x)}=x$. $\mathcal{W}_0(x)$ and $\mathcal{W}_{-1}(x)$ coincide when $x= - e^{-1}$, i.e. $\theta=1$, when $\mathcal{W}_{-1}=\mathcal{W}_0=-1$. This is the limit where the minimum and the maximum coincide and the potential develops a runaway. This implies that we need to consider $\theta<1$. When instead $\theta\to 0$, $\mathcal{W}_{-1}\to -\infty$ and $\mathcal{W}_0\to 0$, implying that in this case $\tau_{\rm max}\to \infty$ and $\tau_{\rm min}\to e^{2/9}\,e^{c/g_s^2}$. This is the opposite limiting case where the uplifting contribution vanishes and we recover just an AdS minimum. Hence a dS minimum can exist only for $0<\theta<1$. Values of $\theta$ close to zero would give rise to an AdS minimum together with a dS maximum, while values of $\theta$ close to unity would uplift the minimum to positive energy. This implies from (\ref{ThetaDef}) that $\alpha$ should be of order $\alpha\simeq (\xi/c)\sqrt{g_s/\tau_{\rm min}}$, as expected by power-counting arguments by staring at \eqref{eq:late-time-pot}. More precisely, we can plug the expression for the minimum \eqref{eq:late-time-extrema-alpha-subbed} in the potential (\ref{eq:late-time-pot}), and impose $V(\tau_{\rm min})>0$, finding the following condition for the existence of a dS minimum:
\begin{equation}
\frac{81 \theta_*}{ 8\left( e^{1 + \mathcal{W}_0(-\theta_*/e)} + 9 \theta_* \right) } > 1\quad\Rightarrow\quad \theta_* \simeq 0.993 < \theta < 1\,.
\end{equation}
This relation can be used to estimate the value of $\alpha$ using (\ref{ThetaDef}). Given that most of the known Calabi-Yau threefolds with $\xi>0$ have $\xi \in (0.1,1.5)$ \cite{Kreuzer:2000xy},\footnote{Explicit global CY examples have shown that the O7-dependent correction to $\xi$ in (\ref{xiDef}) is in general not large enough to change the sign of $\xi$ or to reduce its size significantly (see for example \cite{Cicoli:2017shd,Cicoli:2021dhg}).} we can set in (\ref{ThetaDef}) $\xi= 1$ together with $c=\zeta(3)/\pi^3$ from (\ref{eq:xi-D-sigma-relation}), obtaining:
\begin{equation}
\theta\simeq 1\quad\Leftrightarrow\quad \alpha \simeq  4.78 \sqrt{g_s}\,e^{-\frac{0.02}{g_s^2}}\simeq 0.22\quad\text{for}\quad g_s\simeq 0.1\,,
\end{equation}
which for example would be exactly in the right ballpark if $\alpha$ were given in terms of the 1-loop $\beta$-function coefficient of an $SU(2)$ theory as $\alpha= \beta_0/(8\pi) = 3/(4\pi)\simeq 0.24$. Note that similar values of $\alpha$ can be obtained for different values of $\xi$ and $g_s$ (for example $\xi=0.1$ would require $g_s\sim 0.3$). We therefore conclude that late-time dS minima should naturally exist in the string landscape for appropriate values of the flux quanta which determine $g_s$. The behaviour of the late-time scalar potential for different values of $\theta$ is presented in Fig. \ref{fig:late-time-mins-plot-maintext}. Let us stress that (\ref{eq:late-time-extrema-alpha-subbed}) receives subleading corrections but in App. \ref{sec:Subleading Appendix Late Time} we show that subdominant contributions do not modify our results qualitatively - rather they make our scenario more robust. 

\section{Early-time moduli stabilisation: brane-antibrane inflation}
\label{sec:inflation-alpha}

In Sec. \ref{sec:late-time} we have shown that perturbative corrections to the K\"ahler potential can yield a dS post-inflationary minimum for the volume modulus. In this section we focus instead on brane-antibrane inflation and stabilise again the volume modulus perturbatively in order to avoid the $\eta$-problem.

To do so, we need to consider three fields: the K\"ahler modulus $\tau \equiv (T+\bar{T})$, the inflaton $r$ that controls the distance of the D3-brane from the $\overline{\rm D3}$-brane at the tip of the throat, and a nilpotent superfield $X$ which describes the dynamics of the $\overline{\rm D3}$-brane and realises supersymmetry non-linearly.\footnote{See \cite{Antoniadis:2024hvw} for a recent review and \cite{Guleryuz:2022hiv} for recent studies on non-linear supergravity and cosmology.} In fact, $X$ is subject to the constraint $X^2=0$ that projects out all bosonic degrees of freedom, leaving just the Goldstino in the low-energy EFT.\footnote{Ref. \cite{Burgess:2022nbx} considered instead the case where the inflaton $r$ is the scalar component of a chiral superfield $\Phi$ subject to the constraint $\overline{\mathcal{D}}(X \Bar{\Phi})=0$ which projects out the fermionic and auxiliary fields. In our treatment (see also \cite{Aparicio:2015psl}) we do not need to include the inflaton in the sector which realises supersymmetry non-linearly. In fact, due to the geometrical separation between the D3- and the $\overline{\rm D3}$-brane which realises supersymmetry non-linearly, we expect the inflaton to be the component of a standard chiral superfield which realises supersymmetry linearly.}

As already pointed out in Sec. \ref{Sec2}, after introducing the D3-position radial modulus $r$, the physical Calabi-Yau volume $\sigma = \mathcal{V}^{2/3}$ is not given anymore by $\tau$ but by (see (\ref{ChiralCoord})):
\begin{equation}
\sigma = \tau - \frac16 \left(M_\KK r\right)^2\,.
\label{ChiralCoord2}
\end{equation}
It is this quantity, $\sigma$, and not $\tau$, that is stabilised perturbatively during inflation, guaranteeing the absence of the $\eta$-problem. Note that after the end of inflation $r\to 0$ and $\sigma\to \tau$. The 4D supergravity EFT can be expanded in powers of $X$ as follows \cite{Burgess:2022nbx,Burgess:2021obw,DallAgata:2015zxp,McDonough:2016der,Kallosh:2017wnt}:
\begin{eqnarray}
K&=&-3 \ln \left[f(\sigma)+ (X+\overline{X})\, g (\sigma)  -X \overline{X}h(\sigma)\right], \nonumber \\
W&=&W_0 +X\, W_X (r)\,,
\label{EFT}
\end{eqnarray}
where $K$ does not depend on $(T-\bar{T})$ due to the axionic shift symmetry that is exact at perturbative level. Combining this symmetry with the holomorphy of $W$, one infers that the perturbative superpotential cannot depend on $T$.\footnote{The superpotential could also contain $T$-dependent non-perturbative terms that are often used to freeze the moduli. However we do not consider these terms here as we are working at large volume where such terms are suppressed compared to perturbative corrections if $W_0$ is not tuned to exponentially small values.} We shall also exploit the fact that tree-level string theory features an accidental scale invariance broken by quantum effects that can be organised in an expansion in powers of $1/\sigma$ \cite{Burgess:2020qsc}. Hence we will consider the functions $f(\sigma)$, $g(\sigma)$ and $h(\sigma)$ in the K\"ahler potential as series in $1/\sigma$ with additional powers of $(\ln\sigma)$, as typical of loop corrections. Moreover, the requirement to obtain known tree-level physics at leading order can be used to infer the dominant terms of these functions. In fact, the tree-level expressions of these functions are $f(\sigma)=\sigma$, $g(\sigma)=0$ and $h(\sigma)=1$. The first non-zero contribution to $g(\sigma)$ could hence arise only at loop-level with a potential logarithmic enhancement, giving $g(\sigma) = g_s\ln\sigma$. 

To derive the F-term supergravity scalar potential we take derivatives with respect to the three fields $\tau$, $r$ and $X$, and then we set $X=\overline{X}=0$ and we work in the limit where $r$ is smaller than the size of the extra-dimensions, i.e. $r\ll M_\KK^{-1}$. Hence we find:
\begin{eqnarray}
e^{-K}\,V &=&\left[K^{\overline{X}X}\,W_X^2+W_0 W_X\left(K^{\overline{X}A}K_A+ K^{\overline{A}X} K_{\overline{A}}\right)\right. \nonumber \\
&+&\left.\left. W_0^2\left(K^{\overline{A}B} K_{\overline{A}} K_B-3\right)\right]\right|_{X=0},
\label{VFterm}
\end{eqnarray}
where the index $A$ runs just over $T$ and $X$ since $K_r\simeq 0$ for $r\ll M_\KK^{-1}$, and we assumed without loss of generality that $W_X\in\mathbb{R}$. Using the notation $d y/d\sigma\equiv y'$, the F-term potential (\ref{VFterm}) becomes:
\begin{eqnarray}
V&=& \frac{1}{U}\left[\left(f' W_X-3 g' W_0\right)^2 \right. \nonumber \\
&-&\left. f'' \left(f W_X^2  - 6 g W_X W_0 -9 h W_0^2\right)\right],
\label{GenPot}
\end{eqnarray}
where:
\begin{equation}
U\equiv 3 f^2 \left(2 g f' g' - f g'^2 + f'^2 h - f'' (g^2 + f h)\right).
\end{equation}
As already pointed out, at tree-level we have $f=\sigma$, $g=0$ and $h=1$. In this case one has therefore $f'=1$ and $g'= f''=0$, which implies that (\ref{GenPot}) simply reduces to:
\begin{equation}
V= \frac{W_X(r)^2}{3\sigma^2}\,.  
\label{Vtree}
\end{equation}
As expected, this is the standard $\overline{\rm D3}$-brane uplift contribution if we identify $W_X$ with the warp factor. Introducing also the dependence on $r$ as:
\begin{equation}
W_X (r) = e^{-2\rho}\sqrt{\frac{3}{4\pi}\left(1-\frac{\mathcal{D}_0}{\left(r M_\KK\right)^4}\right)}\,,
\label{WXrho}
\end{equation}
with:
\begin{equation}
\rho\equiv \frac{2\pi K}{3 g_s M}\,,
\end{equation}
the potential (\ref{Vtree}) reproduces the inflationary potential (\ref{VVinf}). Let us stress that the form of $W_X$ is inferred by requiring that the scalar potential term of the DBI $\overline{\rm D3}$-brane action should also be expressible as a standard supergravity F-term potential. However $\sigma$ would be an unstable runaway direction at tree-level. We need therefore to add quantum corrections to fix $\sigma$. In what follows we shall analyse two separate cases: the first with a linear dependence of the K\"ahler potential on the nilpotent superfield $X$, i.e. $g(\sigma)\neq 0$, while the second without a contribution in $K$ which is linear in $X$, i.e. $g(\sigma)=0$.

\subsection{Case with linear contribution}
\label{subsec:linear}

Let us focus on the case where the leading quantum corrections induce a linear dependence of $e^{-K/3}$ on the nilpotent superfield $X$, so that $f=\sigma$, $g=g_s\ln\sigma$ and $h=1$. In this case the relevant quantities are $f'=1$, $f''=0$ and $g'=g_s/\sigma$, giving a potential which is a perfect square:
\begin{eqnarray}
V&=& \frac{1}{3 \sigma^2 }\left(W_X-\frac{3 g_s W_0}{\sigma}\right)^2 \left[1+\frac{g_s^2}{\sigma}\left(2 \ln\sigma - 1\right) \right]^{-1} \nonumber \\
&\simeq& \frac{1}{3 \sigma^2 }\left(W_X-\frac{3 g_s W_0}{\sigma}\right)^2\,.
\label{PerfectSquare}
\end{eqnarray}
This potential admits a Minkowski minimum at:
\begin{equation}
\sigma_{\rm min}=3 g_s \frac{W_0}{W_X} \simeq 2\sqrt{3\pi}\, g_s W_0\, e^{2\rho}\gg 1\quad\text{for}\quad e^{2\rho}\gg 1\,.
\label{SigmaMin}
\end{equation}
The volume field $\sigma$ is now stabilised but inflation cannot be realised since the vacuum energy vanishes. We need therefore to add the leading perturbative correction to $f(\sigma)$ to uplift this vacuum to dS. In Sec. \ref{sec:late-time} we already classified all known perturbative corrections to the type IIB low-energy EFT and argued that the leading ones are those displayed in the scalar potential (\ref{eq:late-time-pot}). To understand which one of these contributions is the most relevant during inflation, let us consider the following argument.

After the end of inflation, the D3-brane annihilates with the $\overline{\rm D3}$-brane and the potential (\ref{PerfectSquare}) disappears. The volume mode has then to relax at the minimum of the potential (\ref{eq:late-time-pot}), instead of running away towards decompactification. To guarantee that the volume reaches the post-inflationary minimum, the inflationary minimum (\ref{SigmaMin}) has therefore to rely at a field value which is smaller than the one at the post-inflationary maximum given by (\ref{eq:late-time-extrema-alpha-subbed}), i.e. $\sigma_{\rm min}<\tau_{\rm max}$. In particular, we shall focus on the case with $\sigma_{\rm min}<\tau_{\rm min}$ where the leading order perturbative correction to $V$ is the $\mathcal{O}(\alpha'^3)$ term in (\ref{eq:late-time-pot}). Hence we consider the case with:
\begin{equation}
f(\sigma)=\sigma+\frac{\xi}{3 g_s^{3/2}\sqrt{\sigma}}\,,\quad g(\sigma)=g_s\ln\sigma\,,\quad h(\sigma)=1\,.
\end{equation}
In this case we obtain a leading order scalar potential of the form:
\begin{equation}
V\simeq \frac{1}{3 \sigma^2}\left( W_X-\frac{3 g_s W_0}{\sigma}\right)^2 + \frac{3\xi W_0^2}{4 g_s^{3/2} \sigma^{9/2}}\,.
\label{SigmaPotInfl}
\end{equation}
At the post-inflationary minimum, the $\xi$-dependent contribution is of the same order of magnitude of the term in (\ref{eq:late-time-pot}) proportional to $\alpha$ that scales as $3\alpha W_0^2/\sigma^4$. Given that the potential (\ref{PerfectSquare}) features the term $3g_s^2 W_0^2/\sigma^4$ which scales in the same way with $W_0$ and $\sigma$, if $\alpha\gtrsim g_s^2$, the $\alpha$-dependent piece would be of comparable size. In turn, the $\xi$-dependent term, that dominates over the $\alpha$-dependent one for field values around the putative inflationary minimum at $\sigma_{\rm min}<\tau_{\rm min}$, would not be a small correction in (\ref{SigmaPotInfl}). On the other hand, it would be the main contribution that would induce a runaway. Hence we need to impose $0<\alpha\ll g_s^2$. In this case the $\xi$-dependent term in (\ref{SigmaPotInfl}) can be a small correction, and so (\ref{SigmaMin}) is still a very good approximation for the location of the minimum which however now becomes dS since, to get a viable post-inflationary minimum, we are focusing on the case where $\xi>0$.

\begin{figure*}[t]
\centering
\includegraphics[scale=1.0]{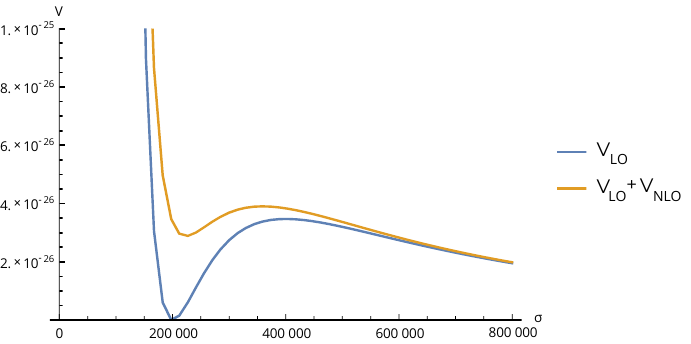}
\caption{Example of an uplifted dS minimum during inflation for the case with linear contribution. The parameter $\alpha$ has been fixed as in (\ref{ThetaDef}) with $\theta=0.994$ to obtain also a late-time minimum. The parameter choices of this example are $g_s =1/15$, $W_0= 1$, $W_X =10^{-6}$, $T_7 = 2 \pi$ and $\xi = 0.1$. The blue curve is the leading order potential (\ref{PerfectSquare}) which features a Minkowksi minimum, while the yellow curve includes the contribution of the next-to-leading order potential that uplifts the vacuum energy to a positive value.}
\label{fig:inflation_comparisonNew}
\end{figure*}

In App. \ref{sec:Appendix Linear} we perform a detailed analysis of the effect of all known higher order quantum corrections, finding that their effect does not alter our conclusions qualitatively. Stable dS inflationary minima can thus be constructed in two steps: finding first a Minkowski minimum of the leading order potential (\ref{PerfectSquare}), and then uplifting it to dS by including subdominant quantum effects. In Fig.~\ref{fig:inflation_comparisonNew} we provide an explicit example of such an inflationary minimum including for completeness, not just the $\xi$-dependent term in (\ref{SigmaPotInfl}), but also additional corrections like the one proportional to $\alpha$ (see (\ref{eq:inflation_VLO_VNLO_VNNLO_defs}) for more details). 

A very good analytical approximation for the value of the vacuum energy can be obtained by substituting (\ref{SigmaMin}) in (\ref{SigmaPotInfl}), obtaining:
\begin{equation}
V(\sigma_{\rm min})\simeq\frac{3\xi W_0^2}{4 g_s^{3/2} \sigma_{\rm min}^{9/2}} \simeq \frac{\xi}{108\sqrt{3} g_s^6} \frac{W_X^{9/2}}{W_0^{5/2}}\,.
\label{Vmin}
\end{equation}
If we now substitute in (\ref{Vmin}) the expression (\ref{WXrho}) of $W_X$ as a function of the inflaton $r = \varphi/\sqrt{T_3}$,  and we expand for large field values, we obtain an inflationary potential of the same form as the power-law general example in (\ref{GenPot0}) with $n=4$:
\begin{equation}
V_{\rm inf}(\varphi) \simeq C_0\left(1-\frac{C_1}{\varphi^4 }\right)\,,
\label{LinerPot}
\end{equation}
where:
\begin{equation}
C_0\equiv \frac{9\xi}{192\sqrt{2} (3\pi)^{9/4} g_s^6} \frac{e^{-9\rho}}{W_0^{5/2}} \quad\text{and}\quad C_1\equiv \frac{9\mathcal{D}_0 T_3^2}{4 M_\KK^4}\,.
\end{equation}
This potential scales as the original brane-antibrane potential (\ref{Vtree}) but with the difference that now $\sigma$ is stabilised perturbatively at $\sigma=\sigma_{\rm min}$. This has two very important implications: ($i$) inflation can occur with a stable volume direction; ($ii$) this model is not plagued by any $\eta$-problem since perturbative corrections to the effective action stabilise $\sigma = \tau- \gamma r^2$, instead of just $\tau$. 

\begin{figure}[t]
\centering
\includegraphics[scale=1.0]{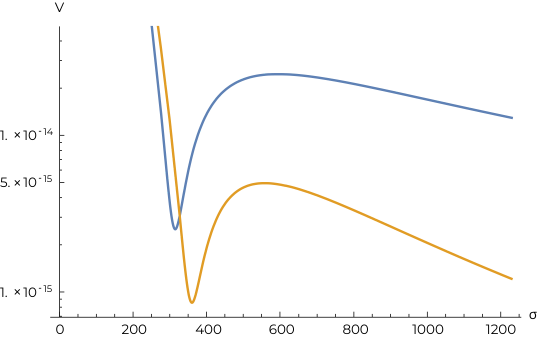}
\caption{The blue curve is the inflationary potential for the case with linear contribution which features a dS minimum with $V_{\rm min}^{\rm inf}\simeq 2.5\times 10^{-15}$, while the yellow curve is the late-time potential with a smaller vacuum energy at $V_{\rm min}\simeq 8\times 10^{-16}$ and a minimum at larger field values. The parameters used to generate this plot are $\xi= g_s=0.1$, $W_0=1$, $W_X = 10^{-3}$ and $T_7 =2\pi$, while $\alpha$ has been fixed as in (\ref{ThetaDef}) with $\theta=0.994$.}
\label{fig:The Linear Term}
\end{figure}

Let us finally stress that in Fig. \ref{fig:inflation_comparisonNew} the parameter $\alpha$ is set at its late-time value (\ref{ThetaDef}) to guarantee the existence of a post-inflationary minimum as well. Fig. \ref{fig:The Linear Term} shows the existence of a dS minimum during inflation and an almost Minkowski vacuum at late times. In the example of Fig. \ref{fig:The Linear Term}, during inflation the volume mode is stabilised around $\sigma_{\rm min}\sim 300$ and after the end of inflation evolves to the new minimum at $\tau_{\rm min}\sim 370$. This implies that the canonically normalised volume mode $\chi$ travels through a distance in field space of order $\Delta \chi = \sqrt{\frac32}\ln\left(\tau_{\rm min}/\sigma_{\rm min}\right)\sim 0.25\,M_p$. In this particular example, this distance turns out to be sub-Planckian but one could easily choose different underlying parameters which would push $\tau_{\rm min}$ to much larger values, obtaining a trans-Planckian distance between the inflationary and the late-time minimum, as envisaged in \cite{Conlon:2022pnx,Apers:2022cyl,Apers:2024ffe}.

\subsection{Case without linear contribution}
\label{subsec: no linear}

Let us now move to the case where $e^{-K/3}$ does not feature any linear dependence on the nilpotent superfield $X$, so that $g(\sigma)=0$. The only effect of the $\overline{\rm D3}$-brane at the tip of the throat is thus just to generate the standard uplifting contribution (\ref{Vtree}).

\begin{figure}[t]
\centering
\includegraphics[scale=0.5]{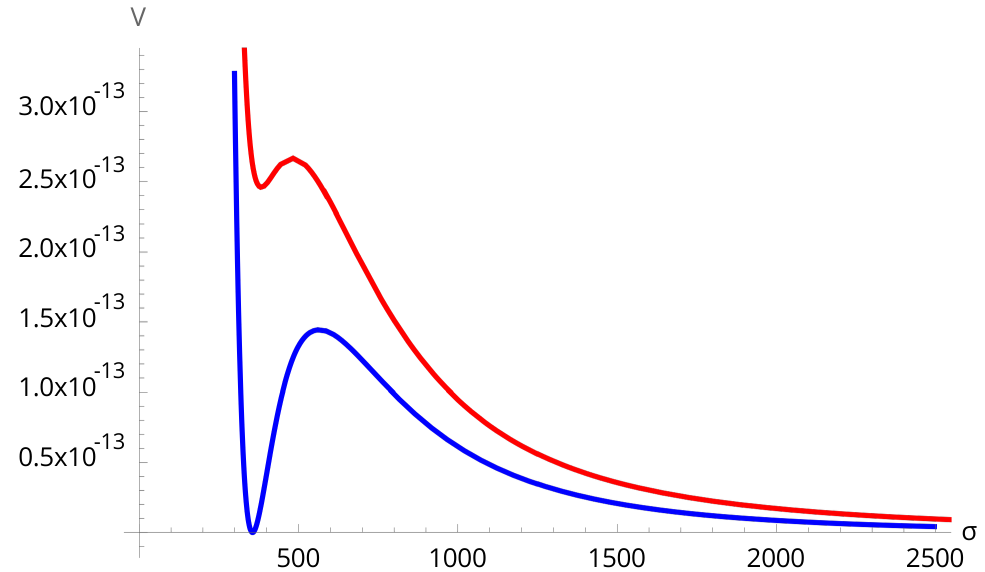}
\caption{The red curve is the inflationary potential for the case without linear contribution which features a dS minimum with $V_{\rm min}^{\rm inf}\simeq 2.5\times 10^{-13}$, while the blue curve is the late-time potential with a much smaller vacuum energy but with a minimum at similar field values. The parameters used to generate this plot are $\xi=g_s =0.1$, $W_0=1$, $W_X^2 = 10^{-7}$ and $T_7 =2\pi$, while $\alpha$ has been fixed as in (\ref{ThetaDef}) with $\theta=0.994$.}
\label{fig:The No-Linear Term}
\end{figure}

To stabilise the volume mode $\sigma$ we need to include quantum corrections to the function $f(\sigma)$ which have already been exploited in Sec. \ref{sec:late-time} to find a late-time minimum. Hence the potential during inflation is the same as (\ref{eq:late-time-pot}) with the addition of the $\overline{\rm D3}$-brane uplifting contribution (\ref{Vtree}), giving:
\begin{equation}
V \simeq\frac{W_X^2}{3\sigma^2}  + 3 W_0^2\left[\frac{\alpha}{\sigma^4} - \frac{\xi\sqrt{g_s}}{4 c \sigma^{9/2}}\left( \ln\sigma-\frac{c}{g_s^2}\right)\right].
\label{eq:inf-pot-no-linear-VNLO}
\end{equation}
As already seen, the terms proportional to $W_0$ in (\ref{eq:inf-pot-no-linear-VNLO}) are all of the same order around the late-time minimum $\tau_{\rm min}$. Hence, if the location of the minimum varies significantly during inflation, only one of these terms would dominate in that region. However the potential would not be rich enough to generate a minimum since it would feature only this term and the one proportional to $W_X$. Thus we conclude that the presence of a stable minimum for $\sigma$ during inflation requires that $W_X$ is small enough, so that the inflationary and late-time minima are very close in field space, i.e. $\tau_{\rm min}\simeq \sigma_{\rm min}$. Note that obtaining a small $W_X$ is not a problem due to the suppression arising from the warp factor, $W_X\simeq e^{-2\rho}\ll 1$. We therefore consider the situation where the first term in (\ref{eq:inf-pot-no-linear-VNLO}) gives just a small correction to the late-time potential (\ref{eq:late-time-pot}). In this case, the minimum is expected to lie still around (\ref{eq:late-time-extrema-alpha-subbed}), and so the inflationary vacuum energy can very well be approximated as:
\begin{equation}
V(\sigma_{\rm min})\simeq\frac{W_X^2}{3 \sigma_{\rm min}^2} 
\simeq\frac{e^{-\frac{2c}{g_s^2}}}{3 \lambda_0^2}\, W_X^2\,.
\label{VminNew}
\end{equation}
Substituting now in (\ref{VminNew}) the expression (\ref{WXrho}) of $W_X$ as a function of the inflaton $r$, we can reproduce the exact form of the original brane-antibrane potential (\ref{VVinf}):
\begin{equation}
V_{\rm inf} (r) =  \frac{\mathcal{C}_0}{\mathcal{V}_{\rm min}^{4/3}}\left[1-\frac{\mathcal{D}_0}{\left(r M_\KK\right)^4}\right]\,,
\end{equation}
but now with the volume mode which is safely stabilised using purely perturbative corrections to the effective action, so avoiding any $\eta$-problem. Fig. \ref{fig:The No-Linear Term} shows the presence of a dS minimum during inflation and an almost Minkowski vacuum at late times. The effect of subleading corrections to the potential (\ref{eq:inf-pot-no-linear-VNLO}) has been analysed in detail in App. \ref{sec:no-linear-appendix} finding that they do not alter our results qualitatively. 

\section{Allowed UV parameter space}
\label{sec:Parameter-Space-Everything}

In Sec. \ref{sec:inflation-alpha} we have shown how to obtain a brane-antibrane inflationary potential with the volume mode fixed by perturbative effects for two cases depending on the presence or absence of a linear dependence of $e^{-K/3}$ on $X$. In both cases the inflationary potential, after substituting the value of $\sigma$ at the minimum, depends on six underlying parameters $(\xi,\alpha,g_s,W_0,K,M)$. In this section we will fix $\xi=0.1$ and $\alpha$ as in (\ref{ThetaDef}) with $\theta=0.994$ to guarantee a late-time minimum. Hence we will restrict our analysis to four UV parameters $(g_s,W_0,K,M)$ and find which regions of the underlying parameter space can be compatible with observations and with an EFT under computational control. 

\begin{itemize}
\item \textbf{Compatibility with data:} We use the Planck data \cite{Planck:2018jri} for observational constraints on the amplitude of density perturbations and the spectral index $n_s$. It can be shown \cite{Burgess:2022nbx} that CMB observations can be matched at horizon exit around $N_e\simeq 56$ e-foldings before the end of inflation where $\varphi_* \simeq 10^{-3}M_p$ and:
\begin{equation}
V_{\rm inf}\simeq 10^{-17}\,M_p^4\,, \quad n_s \simeq 0.970\quad\text{and}\quad r\simeq 2\times 10^{-8}\,,
\end{equation}
with a tiny tensor-to-scalar ratio $r$ that is far from present observational reach. The value of the scalar potential at horizon exit can be used to express $\rho$ in terms of the other UV parameters after writing $V_{\rm inf}$ as (\ref{Vmin}) for the case with linear contribution, and as (\ref{VminNew}) for the case without linear contribution. This reduces the number of independent UV parameters to three.

\item \textbf{Consistency conditions:} The requirement to have computational control over the EFT leads to several consistency conditions:
\begin{enumerate}
\item \emph{D3-brane in the throat:} To have the D3-brane inside the throat of size $R$ but sufficiently far from the $\overline{\rm D3}$-brane localised at $r_0$, we need to impose $r_0\ll r < R$. Expressing $r_0$ and $R$ in terms of the stabilised volume $\mathcal{V}=\sigma_{\rm min}^{3/2}$ and the flux integers $K$ and $M$ respectively as in (\ref{r0}) and (\ref{R}), and writing everything in terms of the canonically normalised inflaton $\varphi$, we find:
\begin{equation}
\frac{27 \,M K}{4(4\pi)^3 \sigma_{\rm min}^3} \left( \sigma_{\rm min}\, e^{-4\rho}\right) \ll \left(\frac{\varphi}{M_p}\right)^4 <
\frac{27\,M K}{4(4\pi)^3\sigma_{\rm min}^3}\,,
\label{eq:throat-bounds}
\end{equation}
which can only be satisfied for sufficiently strong warping, i.e. $\sigma_{\rm min}\, e^{-4\rho}\ll 1$.

\item \emph{Gravitino mass below warped string scale:} To have a controlled EFT in the warped throat, the gravitino mass should be smaller than the warped string scale (\ref{WarpedStringScale}) that controls the size of massive string states on the $\overline{\rm D3}$-brane at the tip of the throat. This turns into the following condition:
\begin{equation}
\frac{m_{3/2}}{M_{s,{\rm warped}}} \simeq  \frac{g_s^{1/4} W_0}{\sqrt{2}\,\sigma_{\rm min}} \,e^{\rho} \ll 1\,,
\end{equation}
which implies:
\begin{equation}
\sigma_{\rm min} \gg \frac{g_s^{1/4} W_0}{\sqrt{2}} \,e^{\rho}\,.
\end{equation}

\item \emph{Dilute flux approximation under control:} Very large throats at not-too-large-volumes can lead to a failure of the dilute flux approximation, thus modifying the computation of the 4D Planck mass and yielding corrections to the Einstein-Hilbert term \cite{Gao:2022fdi}. Even though this does not need to be problematic for our proposed inflation mechanism (since the modified corrections could still be used for uplifting to the correct vacuum energy), we will be conservative and impose:
\begin{equation}
\sigma_{\rm min} > M K\,.
\end{equation}
This condition should suppress other unknown corrections and avoid a potential singular-bulk problem~\cite{Gao:2020xqh}, related to the throat `not fitting' within the bulk~\cite{Carta:2019rhx}.

\item \emph{Curvature corrections under control:} The tension of the $\overline{\rm D3}$-brane at the tip of the throat can receive curvature corrections \cite{Junghans:2022exo,Junghans:2022kxg} which have been recently studied in \cite{Hebecker:2022zme,Schreyer:2022len,Schreyer:2024pml} in the context of the metastability of the KPV setup \cite{Kachru:2002gs} finding a potential reduction of the tension of the $\overline{\rm D3}$-brane without the requirement of large fluxes (and so potentially reducing the problems mentioned above). For our present discussion, we simply require that these curvature corrections are small by imposing the condition:
\begin{equation}
g_s M > 1\,.
\end{equation} 

\item \emph{No destabilisation of the conifold modulus:} It has been shown in \cite{Bena:2018fqc} that the backreaction of the antibrane can destabilise the conifold modulus unless:
\begin{equation}
g_s M^2\geq 46\,.
\label{MildBound}
\end{equation}
This instability was reassessed and challenged in~\cite{Lust:2022xoq} where the authors argued that the stabilisation of the conifold modulus is not a problematic issue as long as $g_s M^2\gg N_{\overline{D3}}$. Assessing the correctness of this bound is beyond the scope of our paper. To be conservative, we simply impose it even if it shrinks a bit the allowed UV parameter space of our models.
\end{enumerate}
\end{itemize}

\begin{figure}[h!]
\centering
\includegraphics[scale=0.7]{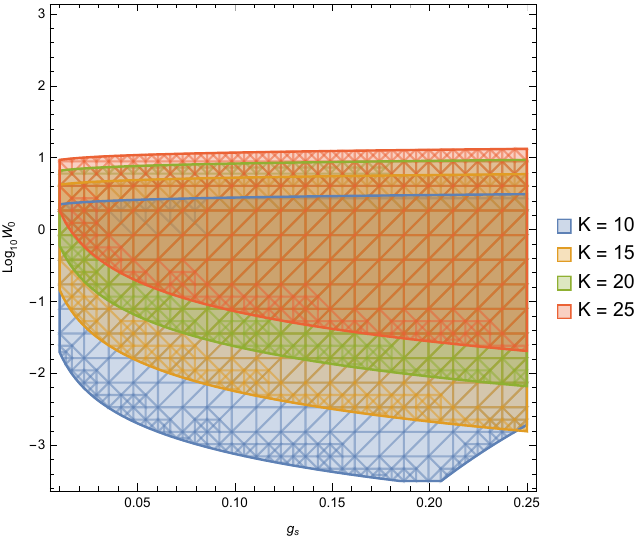}
\caption{Allowed UV parameter space for the case with linear contribution. The parameters used to generate this plot are $\xi= 0.1$ and $T_7 =2\pi$, while $\alpha$ has been fixed as in (\ref{ThetaDef}) with $\theta=0.994$.}
\label{fig:linear-inf-region-plot} 
\end{figure}

\begin{figure}[h!]
\centering
\includegraphics[scale=0.7]{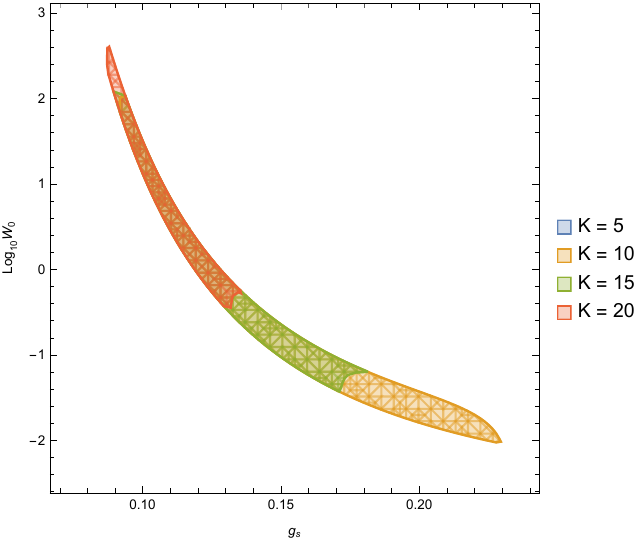}
\caption{Allowed UV parameter space for the case without linear contribution. The parameters used to make this plot are $\xi= 0.1$ and $T_7 =2\pi$, while $\alpha$ has been fixed as in (\ref{ThetaDef}) with $\theta=0.994$.}
\label{fig:no-linear-inf-region-plot}
\end{figure}

Clearly, satisfying all these conditions with full parametric control is not possible. However, in both cases there is an allowed region of the underlying UV parameter space where all constraints can be satisfied with numerical control. As argued above, after imposing compatibility with data, we are leftover with three independent UV parameters which we choose to be $(W_0,g_s,K)$. We then plot the region in the $(W_0,g_s)$ parameter space that satisfies all consistency conditions for different values of $K$. Fig. \ref{fig:linear-inf-region-plot} shows our results for the case with linear contribution, while Fig. \ref{fig:no-linear-inf-region-plot} refers to the case without linear contribution. 

Note that the allowed region of the UV parameter space shrinks as the flux $K$ increases and that the case with linear contribution features a larger allowed parameter space. Let us conclude by showing in Tab. \ref{tab:Numeric_example} an illustrative numerical example for both cases where all constraints are satisfied.

\begin{table}[h!]
\centering
\begin{tabular}{c|c|c|c||c|c}
Cases & $K$ & $ g_s  $ & $W_0$ & $M$ & $\sigma_{\rm min}$  \\
\hline
With linear term & $10$ & $0.15 $ & $\simeq 0.24$ & $29$ &  $\simeq 3360$ \\  \hline
Without linear term & $10$ & $0.15 $ & $ \simeq 0.16$ & $23$ &  $\simeq 670$ \\  \hline
\end{tabular}
\caption{Benchmark examples which satisfy all constraints for the parameters $(W_0,g_s,K)$ and the resulting values of $M$ and $\sigma_{\rm min}$.}
\label{tab:Numeric_example}
\end{table}

Albeit a bit large, the tadpoles arising from the flux integers shown in Tab. \ref{tab:Numeric_example}, $KM=290$ and $KM=230$ for the two cases respectively, could in principle be cancelled in realistic models~\cite{Kreuzer:2000xy}. Let us finally point out that higher order $\alpha'$ corrections might require a very large value of $g_s M^2$ to avoid the KPV decay of the $\overline{\rm D3}$-brane \cite{Schreyer:2022len,Schreyer:2024pml}. Even if the corresponding decay channel is not well understood for a  single $\overline{\rm D3}$-brane, it is easy to check that imposing $g_s M^2\geq 150$, instead of $g_s M^2\geq 46$ as in (\ref{MildBound}), there is still a large allowed region of underlying parameter space. For instance, in the linear case, $g_s=0.05$, $K=10$ and $W_0=0.1$ would give $M=70$, and so $g_sM^2=245$. The only potential problem for this case would be a rather large tadpole, $KM=700$. We leave however a concrete realisation of our model for future work.

\section{Conclusions}

In this paper, expanding on a proposal in \cite{Burgess:2022nbx}, we derived the scalar potential of brane-antibrane inflation within a low-energy EFT of type IIB string compactifications where the volume mode is stabilised by perturbative effects and inflation is driven by the motion of a D3-brane that is attracted by an $\overline{\rm D3}$-brane at the tip of a warped throat. The key features of our construction are the following ones:
\begin{itemize}
\item We focused on the brane-antibrane inflationary model proposed in \cite{Burgess:2001fx,Dvali:2001fw}.

\item We followed the approach of \cite{Kachru:2003sx} which introduced warping to obtain a flat potential that can drive inflation with an interbrane separation that is correctly smaller than the size of the extra dimensions.

\item Contrary to \cite{Kachru:2003sx}, which proposed to fix the volume mode with non-perturbative effects, we stabilised the volume exploiting leading known $\alpha'$ and $g_s$ corrections to the type IIB effective action. In this way we managed to avoid the $\eta$-problem and derived for the first time the inverse power-law potential originally proposed in \cite{Burgess:2001fx,Dvali:2001fw}, instead of the tuned inflection point inflation model of \cite{Baumann:2007ah} that is now in conflict with observations.

\item We described the Coulomb interaction of the D3-$\overline{\rm D3}$-pair in a manifestly supersymmetric EFT thanks to the introduction of a nilpotent superfield $X$. 

\item We considered the most general expansion of $e^{-K/3}$ in powers of $X$. In the presence of a linear dependence of $e^{-K/3}$ on $X$, we discovered a new perfect-square structure of the scalar potential which allows to fix the volume mode during inflation at a Minkowski minimum which is uplifted to dS by subdominant $\mathcal{O}(\alpha'^3)$ corrections. If instead $e^{-K/3}$ does not feature any linear dependence on the nilpotent superfield, we showed that the volume mode can still be fixed at a dS vacuum by exploiting logarithmic moduli redefinitions. 

\item After imposing compatibility with observations and consistency conditions over the EFT, we found large regions of the underlying UV parameter space where brane-antibrane inflation can be successfully realised with enough e-foldings, the right amplitude and spectral index of scalar fluctuations and a tensor-to-scalar ratio far below current observational bounds. 
\end{itemize}

Let us stress that our new framework for brane-antibrane inflation based on perturbative moduli stabilisation avoids the problems faced by the original proposal but shares its interesting aspects. In particular, the end of inflation and reheating can have a purely stringy nature, contrary to many other string inflationary models where this epoch can instead be described within a low-energy supergravity perspective. In fact, brane-antibrane inflation typically ends as in hybrid inflation scenarios with the open string inflaton that becomes the tachyonic waterfall field at a critical interbrane separation.
The arising of a tachyon \cite{Sen:1998sm,Sen:2002nu,Sen:2002in} captures the process of brane-antibrane annihilation whose endpoint is a gas of highly excited strings. These string modes eventually decay giving rise to successful reheating \cite{Kofman:2005yz}. Before decaying, these strings reach thermal equilibrium (in the presence of branes) in a Hagedorn phase \cite{Frey:2005jk,Frey:2023khe} which leads to the emission of gravitational waves peaked in the GHz band \cite{Frey:2024jqy} (see~\cite{Aggarwal:2020olq} for a review).
If the strings are instead stable, they behave as stringy cosmic strings \cite{Sarangi:2002yt,Copeland:2003bj,Majumdar:2005qc,Conlon:2024uob} that may be tested experimentally, independently of CMB tests. Additionally, hybrid inflation models have been found to generate primordial black holes (see for instance  \cite{LISACosmologyWorkingGroup:2023njw}). Studying in detail the dynamics or reheating and exploring the possibility to generate primordial black holes is very interesting but it is beyond the scope of our paper, and so we leave it for further studies. 

Another important question is the one of realising a concrete Calabi-Yau orientifold compactification with explicit brane setup and tadpole cancellation which features an $\overline{\rm D3}$-brane at the tip of a warped throat together with 3-form fluxes that stabilise the dilaton and the complex structure moduli in such a way to reproduce values of $g_s$ and $W_0$ that lie within the allowed parameter space of our construction. Interesting recent works in this direction are \cite{Kallosh:2015nia,Crino:2020qwk,AbdusSalam:2022krp,McAllister:2024lnt}. We plan to exploit in the future the results of these papers to address the crucial issue of embedding brane-antibrane inflation in globally consistent Calabi-Yau compactifications.

\section*{Acknowledgements}

We would like to thank Sebastian Cespedes, Joseph Conlon and Simon Schreyer for stimulating discussions. This article is based upon work from COST Action COSMIC WISPers CA21106, supported by COST (European Cooperation in Science and Technology).
The work of CH has been partially supported by Cambridge Trust and NUI. The work of FQ and GV has been partially supported by STFC consolidated grants ST/P000681/1, ST/T000694/1. The work of M. R-H has been partially supported by Cambridge Trust, CONAHCYT, DAMTP and St. Edmund's College. ARK is supported by Czech Science Foundation GAČR grant ``Dualities and higher derivatives" (GA23-06498S). ARK would like to thank L. Wulff, M. Majumdar, S. Zacarias, R. Mahanta, M. M. Faruk and S. T. Jahan for helpful discussions and Faculty of Science at Masaryk University for supporting travel related to the work. M. R-H thanks the CERN theory department for its hospitality.
GV thanks the University of Bologna and the CERN theory department for hospitality.

\appendix
\section{Brane-antibrane inflation in warped compactifications}
\label{sec:coulomb-potential}

Our starting point is a compactification of type IIB string theory on a Calabi-Yau orientifold in the presence of fluxes. The metric of the 10D space factorises and is given by:
\begin{eqnarray}
d s^2&=&\left(1+\frac{e^{-4 A(z)}}{\mathcal{V}^{2/3}}\right)^{-1 / 2} \eta_{\mu\nu} dx^\mu dx^\nu \nonumber \\
&+&\left(1+\frac{e^{-4 A(z)}}{\mathcal{V}^{2/3}}\right)^{1/2} g_{\alpha\bar{\beta}}dz^\alpha\, d\bar{z}^{\bar{\beta}}\,.
\label{eq:warped-metric}
\end{eqnarray}
Unwarped regions are characterised by $e^{-4 A}\ll \mathcal{V}^{2/3}$, while highly warped throats feature $e^{-4 A}\gg \mathcal{V}^{2/3}$ and are well described by the deformed conifold geometry (the famous Klebanov-Strassler solution \cite{Klebanov:2000hb}) with an internal metric of the form:
\begin{equation}
d s_6^2 \simeq \frac{e^{-2 A(r)}}{\mathcal{V}^{1/3}} \left(dr^2+ r^2 ds^2_{T^{1,1}}\right)\,.
\label{ConifoldMetric}
\end{equation}
Here $r$ is the radial coordinate and $T^{1,1}$ is the Sasaki-Einstein manifold which is an $S^2$ fibration over an $S^3$. The warp factor can be written in terms of the size of the throat $R$ as:
\begin{equation}
\frac{e^{-A(r)}}{\mathcal{V}^{1/6}}\simeq \frac{R}{r}\,,    
\end{equation}
and takes its maximal value at the tip of the throat where the conifold singularity is resolved by an $S^3$ of size $r_0$ \cite{Giddings:2001yu}:
\begin{equation}
e^{-4 A(r_0)} \simeq \left(\frac{R}{r_0}\right)^4 \mathcal{V}^{2/3} \simeq e^{\frac{8\pi K}{3 g_s M}}\equiv e^{4 \rho}\,.
\label{r0}
\end{equation}
Here $K$ and $M$ are respectively the quantised fluxes $F_3$ and $H_3$ wrapping the $S^3$ at the tip (the \textit{A}-cycle) and its Poincaré dual (the \textit{B}-cycle):
\begin{equation}
\frac{1}{\ell_s^2}\int_A {F_3}=M \qquad\text{and}\qquad \frac{1}{\ell_s^2}\int_B {H_3}=K\,.
\end{equation}
At this level, we are free to choose $M$ and $K$ at will, provided the sourced D3-brane tadpole $N\equiv MK$ is cancelled in a global scenario. In this paper we neglected this issue while trying to keep the tadpole as small as possible and consistent with further EFT constraints described in Sec.~\ref{sec:Parameter-Space-Everything}, with the hope that a realistic compactification can cancel it. 
Note moreover that the size of the throat is given in terms of the flux D3-charge $N$ as:
\begin{equation}
R^4\equiv \frac{27\pi}{4} g_s N (\alpha')^2 = \frac{27}{8 (2\pi)^3} \frac{g_s M K}{M_s^4}\,.
\label{R}
\end{equation}

Dimensional reduction determines the string scale $M_s$ in terms of the 4D Planck scale $M_p$ as:
\begin{equation}
M_s =\frac{g_s^{1/4}}{\sqrt{4\pi \mathcal{V}}}\left(1+\frac{e^{-4A(r)}}{\mathcal{V}^{2/3}}\right)^{-1/4} M_p\,,
\end{equation}
which in the unwarped region reduces to (\ref{StringScale}), while in the highly warped region becomes:
\begin{equation}
M_{s,{\rm warped}} =\frac{g_s^{1/4}}{\sqrt{4\pi} \mathcal{V}^{1/3}}\,e^{-\rho} M_p\,.
\label{WarpedStringScale}
\end{equation}
The warped throat plays a prominent role in brane-antibrane inflation because it allows to tune the effective brane tension, and thus flattens the Coulomb potential felt by a brane in the presence of an antibrane. The starting point to study brane dynamics is the worldvolume action of a D3-brane:
\begin{equation}
S_{\rm D3} = S_\DBI + S_\CS\,.
\end{equation}
For our current purposes, all we need is the DBI part:
\begin{equation}
S_\DBI = T_3 \int d^4 x \sqrt{-\operatorname{det} \left(\gamma_{a b}+B_{ab}+2 \pi \alpha^{\prime} F_{a b}\right)}\,,
\end{equation}
where $\gamma_{ab}$ is the pullback of the ambient space metric, $B_{ab}$ is the pullback of the Neveu-Schwarz two-form, and $F_{ab}$ are gauge fluxes on the worldvolume. Inflation is achieved due to the attractive nature of the interbrane potential that appears when one considers the backreaction of a brane in the antibrane action. Before introducing backreaction, we consider the dynamics of branes/antibranes in a warped background. To do so, let us consider a setup with no $B$-field or worldvolume fluxes. Taking an expansion in powers of derivatives and keeping only renormalisable terms, we find:
\begin{eqnarray}
\label{DBID3}
S_\DBI &=&T_3\int{d^4x\,\sqrt{-\operatorname{det}\left(\Tilde{\gamma}_{ab}+\Tilde{g}_{\alpha \bar\alpha}
\partial_a z^\alpha\, \partial_b \bar{z}^{\bar\alpha}\right)}} \\
&=& T_3\int{d^4x\,\left(1+\frac{e^{-4A(r)}}{\mathcal{V}^{2/3}}\right)^{-1}} \nonumber \\
&+&  \int{d^4x\,\left(
\frac12 g_{\alpha \bar \alpha}\partial_\mu (\sqrt{T_3}z^\alpha)\partial^\mu (\sqrt{T_3} \bar{z}^{\bar\alpha})+ \mathcal{O}(\partial z)^4\right)}, \nonumber
\end{eqnarray}
where in the last step we used the unwarped metric $g_{\alpha \bar\alpha}$. The renormalisable part of the Lagrangian thus consists in a vacuum energy and the kinetic terms for the brane position moduli. In the case of D3-branes, the vacuum energy is cancelled by the Chern-Simons part of the action, but for $\overline{\rm D3}$-branes it is doubled. Thus the antibrane position moduli feel a potential that it minimised at the bottom of the throat where the warp factor $e^{-4A}$ takes its maximal value. 

To express the kinetic Lagrangian in terms of the canonically normalised inflaton $\varphi$, let us first recall that the deformed conifold is defined as the locus in $\mathbb{C}^4$ with:
\begin{equation}
\sum_{A=1}^4 z_A^2=\epsilon^2 \, , \qquad \epsilon>0\, .
\end{equation}
Note that the deformation parameter $\epsilon$ measures the size of the $S^3$ at the tip of the throat, and so $r_0 \simeq \epsilon^{2/3}$. The K\"ahler potential for the D3-position moduli sufficiently far away from the tip of the warped throat is very well approximated by the one of the singular conifold with $\epsilon=0$ \cite{Candelas:1989js}:
\begin{eqnarray}
k(z_\alpha, \bar{z}_{\bar{\alpha}})&=&\frac{3}{2}\left(\sum_{A=1}^4 |z_A|^2\right)^{2/3} \nonumber \\
&=& \frac32 \left(\sum_{\alpha=1}^3 |z_\alpha|^2+\left|\sum_{\alpha=1}^3 z_\alpha^2\right|\right)^{2/3}\,.
\end{eqnarray}
The metric $g_{\alpha\bar{\beta}}$ appearing in (\ref{DBID3}) is therefore given by:
\begin{equation}
g_{\alpha\bar{\alpha}} = \frac{\partial^2 k(z_\alpha, \bar{z}_{\bar{\alpha}})}{\partial z_\alpha \partial \bar{z}_{\bar{\alpha}}}\,.
\label{ConifMetric}
\end{equation}
The inflaton in brane-antibrane inflation is the radial distance $r$ between the D3-brane and the $\overline{\rm D3}$-brane at the tip of the throat, and corresponds to $z \sim \bar{z} \sim r^{3/2}$. It is then straightforward to use (\ref{ConifMetric}) to find $g_{\alpha\bar{\alpha}} \sim r^{-1}$. Plugging this result back in the kinetic Lagrangian in (\ref{DBID3}), one finds $\varphi= \sqrt{T_3}\, r$. Note that a proper supergravity embedding requires the inclusion of the volume mode in a tree-level K\"ahler potential of the form:
\begin{equation}
K = -3\ln\left[T+\bar{T} -\gamma k(z_\alpha,\bar{z}_{\bar{\alpha}})\right].   
\end{equation}
It is easy to compute the kinetic terms for the radial distance $r$ and to find $\gamma \sim \langle (T+\bar{T}) \rangle\,T_3$ in order to reproduce the kinetic terms obtained from the DBI action \cite{Baumann:2007ah}.

Let us now discuss the origin of the Coulomb potential which arises when one introduces the backreaction of the D3-brane (located at $r$) in the background metric, yielding a perturbation to the warp factor given by a harmonic function~\cite{Giddings:2001yu,Baumann:2006th}:
\begin{equation}
1+\frac{e^{-4 A}}{\mathcal{V}^{2/3}} \to 1+\frac{e^{-4 A}}{\mathcal{V}^{2/3}}+\delta h(r)\,,
\end{equation}
with:
\begin{equation}
-\nabla^2_{\tilde{r}} \delta h(\tilde{r};r)=\left(\frac{2\pi}{T_3}\right)\frac{ \delta^{(6)}(\tilde{r}-r)}{\sqrt{g(\tilde{r})}}\,.
\end{equation}
This is an eigenvalue problem which can be solved if the metric of the internal space is known. In the conifold case the zero mode of the Laplacian in the angular directions gives the profile $\delta h(r)=\beta/(T_3 r^4)$, with $\beta=27/(32\pi^2)$ \cite{Kachru:2003sx}, while other modes in the multipole decomposition give higher orders in $1/r$. Including this leading effect in the DBI action for the antibrane in the perturbed background, one finds:
\begin{equation}
S_\DBI =
\int{d^4x\,T_3\left(1+\frac{e^{-4 A(r)}}{\mathcal{V}^{2/3}}+\frac{\beta}{T_3\, r^4}\right)^{-1}}\,.
\end{equation}
Adding the Chern-Simons contribution that doubles the $\overline{\rm D3}$-brane tension, we obtain the following potential depending on the strength of the warping:
\begin{equation}
V =\begin{cases}
2T_3 \left(1-\frac{\beta}{T_3\,r^4}\right)\,, \qquad\qquad \qquad \qquad \,\, e^{-4 A}\ll \mathcal{V}^{2/3}\\
2T_3\,\mathcal{V}^{2/3}\,e^{-4 \rho}\left(1-\frac{\beta\, \mathcal{V}^{2/3}\,e^{-4 \rho}}{T_3\,r^4}\right)\,, \qquad e^{-4 A}\gg \mathcal{V}^{2/3} \,.
\end{cases}
\end{equation}
Note that in the strongly warped region the tension of the antibrane is warped down. This is a well understood fact and forms the basis of the antibrane uplift. The idea is that the positive energy from the antibrane can be tuned appropriately by choosing $\rho$. Moreover, the $r$-dependent term can be made small enough to sustain slow-roll inflation, which is generically not the case in the absence of warping~\cite{Burgess:2001fx}.

\section{Late-time potential with subleading terms}
\label{sec:Subleading Appendix Late Time}

\begin{figure*}[h]
\centering
\includegraphics[scale=1]{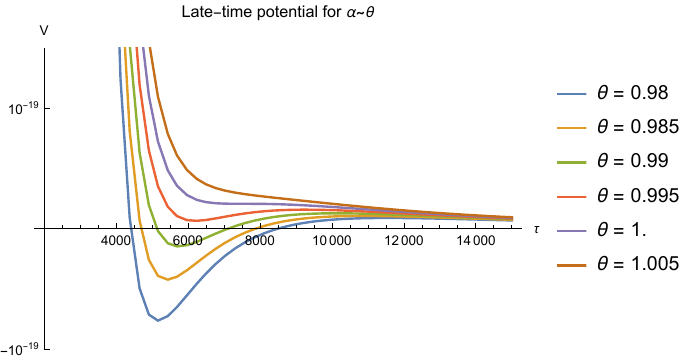}
\caption{Late-time potential with leading order correction for different values of the uplifting parameter $\theta$ setting $\xi=g_s =0.1$ and $T_7=2\pi$. By increasing $\theta$ we move from AdS to dS and finally to a runaway. In particular the red curve shows a dS minimum at $\tau_{\rm min} \sim 6000$ for an appropriate choice of $\theta$.}
\label{fig:late-time-mins-plot-appendix}
\end{figure*}

In this section we study the effect of the leading perturbative correction which has been neglected in Sec. \ref{sec:late-time}. 
We therefore consider the following K\"ahler potential:
\begin{eqnarray} 
K &\simeq&   -3\ln \left[\tau-\alpha\ln \tau \right.\nonumber \\
&+&\left. \frac{\xi}{ 3 c g_s^{3/2}\sqrt{\tau}}\left(c -  g_s^2 \ln\tau\left(1-\frac{1}{3\pi \ln\tau}\right)\right) \right],
\label{eq:Kpot-full-Appendix}
\end{eqnarray}
where we have ignored terms proportional to the tiny constant $c_3\sim 10^{-4}$, as well as loop corrections proportional to $c_1$ and $c_4$ since they would induce a contribution to $V$ only at next-to-next-to-leading order in the $\tau^{-1}$ expansion due to the extended no-scale cancellation. Moreover we did not include subdominant terms arising from the expansion in powers of the small quantity $\alpha=\beta_0/(8\pi)\ll 1$. Note that the K\"ahler potential (\ref{eq:Kpot-full-Appendix}) differs from the leading $K$ (\ref{eq:kahler-pot}) by a correction that scales as $1/(3\pi\ln\tau) \sim g_s^2/(3c\pi)\ll 1$ for $g_s\ll 1$, where we have substituted the extremisation condition (\ref{eq:late-time-extrema-alpha-subbed}). The scalar potential generated by (\ref{eq:Kpot-full-Appendix}) together with $W=W_0$ reads:
\begin{equation}
\frac{V}{3 W_0^2} \simeq\frac{\alpha}{\tau^4} -\frac{\xi\sqrt{g_s}}{4 c\tau^{9/2}}\left(\ln\tau-\frac{c}{g_s^2}-b\right)+\mathcal{O}\left(\frac{1}{\tau^5}\right),
\label{CorrPot}
\end{equation}
with:
\begin{equation}
b\equiv \frac83+\frac{1}{3\pi}\simeq 2.8\,.    
\end{equation}
Note that this potential reproduces the leading one, given by (\ref{eq:late-time-extrema-alpha-subbed}), for $b=0$. The new minimum is at:
\begin{equation}
\tau_{\rm min, new}  = \lambda_0\, e^b \,e^{\frac{c}{g_s^2}} = e^b\,\tau_{\rm min}\,,
\end{equation}
with:
\begin{equation}
\theta_{\rm new} \equiv \left(\frac{16 c\,e^{\frac{10}{9}}}{9 \xi \sqrt{g_s}}\right) \alpha\,e^{\frac{b}{2}}  \, e^{\frac{c}{2 g_s^2}} = e^{\frac{b}{2}}  \,\theta\,.    
\end{equation}
Note that these expressions reduce to (\ref{eq:late-time-extrema-alpha-subbed}) and (\ref{ThetaDef}) for $b=0$. The inclusion of the correction proportional to $b$ does not change the scaling with $g_s$ but it helps to find a larger value of $\tau$ at the minimum, increasing the control over additional corrections. As in the main text, the requirement to obtain a dS vacuum is still $0.993\lesssim\theta_{\rm new}<1$. The corrected potential (\ref{CorrPot}) is plotted in Fig. \ref{fig:late-time-mins-plot-appendix} for different values of $\theta_{\rm new}$ (renamed as $\theta$ in the plot). Comparing Fig. \ref{fig:late-time-mins-plot-appendix} with Fig. \ref{fig:late-time-mins-plot-maintext}, we can indeed see that the physics is unaltered and the only effect of the correction is to shift the value of $\tau$ at the minimum to larger values. 

\section{Early-time potential with subleading terms}
\label{AppC}

\subsection{Case with linear contribution}
\label{sec:Appendix Linear}

\begin{figure}[t]
\centering
\includegraphics[scale=1.0]{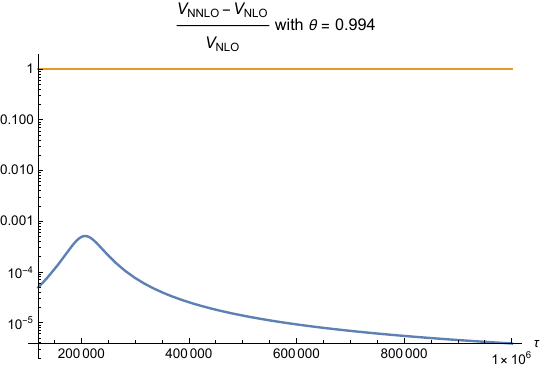}
\caption{Check that the subleading potential $V_\NNLO$ that has not been included in the plot of Fig.~\ref{fig:inflation_comparisonNew} is indeed always negligible.}
\label{fig:inflation_comparison}
\end{figure}

Let us illustrate the robustness of our scenario by including all dangerous known corrections. This is important because, as we have seen, if these corrections are large, the minimum becomes a runaway direction, spoiling inflation. The total potential during inflation can be generically written as:
\begin{equation}
V =\mathcal{A}\,\frac{W_X^2}{\sigma^2}+\mathcal{B}\,\frac{W_0 W_X}{ \sigma^3}+\mathcal{C}\,\frac{W_0^2}{\sigma^4}\,,
\end{equation}
where (ignoring higher order corrections):
\begin{eqnarray}
\mathcal{A} &=& \frac{1}{3} + \frac{g_s^2}{3\sigma}\left(
1 -2 \ln \sigma+ 3  (\ln\sigma )^2 + \frac{2\alpha}{g_s^2} \ln \sigma \right) \nonumber \\
\mathcal{B}&=&- 2 g_s-\frac{2 g_s^3}{\sigma} \left[1 -2 \ln \sigma + 3  (\ln\sigma  )^2  + \frac{\alpha}{g_s^2}\left(\ln\sigma   +2 \right)\right] \nonumber \\
\mathcal{C} &=& 3 \left(g_s^2 + \alpha\right) + 
\frac{3\xi}{4 g_s^{3/2} \sqrt{\sigma }} \left[1 - \frac{g_s^2}{c} \left( \ln \sigma-b \right)\right] 
\label{eq:inflation-terms} \\
&+&\frac{3 g_s^4}{\sigma} \left[1+3 (\ln \sigma)^2+\frac{4\alpha}{g_s^2} +\frac{1}{g_s^4} \left(2 \ln \sigma\left(\alpha^4 -g_s^4 \right)+3 \alpha^2 \right)\right]  \nonumber
\end{eqnarray}
As already stressed in Sec. \ref{subsec:linear}, a stable inflationary minimum can exist if $0<\alpha\ll g_s^2$. In this limit, the scalar potential can therefore be organised as the sum of leading order (LO), next-to-leading order (NLO) and next-to-next-to-leading order (NNLO) contributions as:
\begin{eqnarray}
V_\LO  &=&
\frac{1}{3 \sigma^2}\left( W_X-\frac{3 g_s W_0}{\sigma}\right)^2 \nonumber \\
V_\NLO &=& \frac{3 \xi W_0^2 }{4 g_s^{3/2} \sigma^{9/2}}\left[1 - \frac{g_s^2}{c}\left(\ln\sigma-b \right)\right]+\frac{3 W_0^2 \alpha }{\sigma^4} 
\label{eq:inflation_VLO_VNLO_VNNLO_defs} \\
V_\NNLO &\simeq& \left(\frac{W_X}{W_0}\right)^a\frac{W_0^2}{\sigma^b} \quad\text{with}\quad a+b = 5\,.
\nonumber
\end{eqnarray}
Viable dS minima during inflation can be found by searching first for a Minkowski minimum of $V_\LO$, and then uplifting it to dS by including $V_\NLO$ with $\alpha$ set at its late-time value \eqref{ThetaDef}. This is the procedure followed to obtain the minimum shown in Fig.~\ref{fig:inflation_comparisonNew}. In Fig.~\ref{fig:inflation_comparison} we check that the $V_\NNLO$ contribution that has been neglected in Fig.~\ref{fig:inflation_comparisonNew} is indeed subdominant. 

\begin{figure}[h]
    \centering
    \begin{subfigure}[b]{0.45\textwidth}
       \centering
       \includegraphics[width=\textwidth]{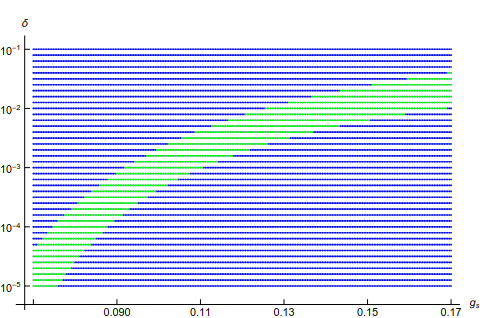}
       \caption{$\xi=0.1$}
       \label{fig:linear_inf_scan1}
    \end{subfigure}
    \hfill
    \begin{subfigure}[b]{0.45\textwidth}
       \centering
       \includegraphics[width=\textwidth]{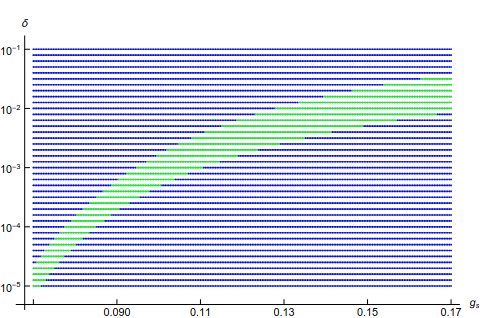}
       \caption{$\xi=1.3$}
       \label{fig:linear_inf_scan4}
   \end{subfigure}
\caption{The green regions of the above parameter spaces (with $\delta\equiv W_X/W_0$) give a stable dS minimum for inflation, while in the blue regions the uplift becomes too large, leading to runaway potentials. This scan was done for the case with linear contribution setting $W_0=1$, $T_7=2\pi$ and $\alpha$ as in (\ref{ThetaDef}) with $\theta=0.994$, implying that in this setup we have both an inflationary and a late-time minimum.}
\label{fig:linear_inf_scan}
\end{figure}

Hence we see that it is possible to construct a dS inflationary minimum for given parameter choices that agree with the late-time minimisation constraints of Sec. \ref{sec:late-time}, requiring only a tuning of $W_X$. To demonstrate the generality of this construction, we performed a numerical scan over the parameters in our theory $\{\xi,g_s,W_X\}$ and found a large region of parameter space where these inflationary minima exist, as shown in Fig.~\ref{fig:linear_inf_scan}.\footnote{As in the Kreuzer-Skarke database, we considered the CY Euler number in the range $\xi \in (0.1,1.5)$.} Fig.~\ref{fig:linear_inf_scan} shows the existence of a suitably uplifted dS inflationary minimum for a given value of $W_X$, provided $g_s$ does not become too small, in which case the potential develops a runaway. Moreover, Fig. \ref{fig:linear_inf_scan} includes just the effect of $V_\NLO$. We have however checked that our results are rather robust since the inclusion of $V_\NNLO$ makes the allowed region of parameter space even a bit larger. 

\subsection{Case without linear contribution}
\label{sec:no-linear-appendix}

\begin{figure}[h]
    \centering
    \begin{subfigure}[b]{0.45\textwidth}
        \centering
          \includegraphics[width=\textwidth]{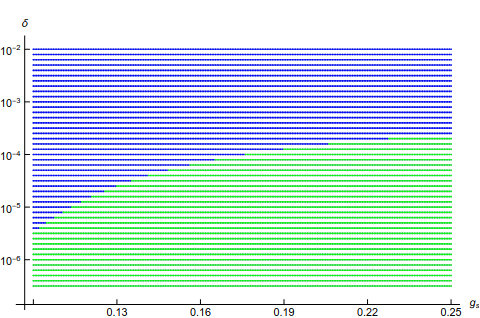}
        \caption{$\xi=0.1$}
        \label{fig:no_linear_inf_scan_naive-A1}
    \end{subfigure}
    \hfill
    \begin{subfigure}[b]{0.45\textwidth}
        \centering
        \includegraphics[width=\textwidth]{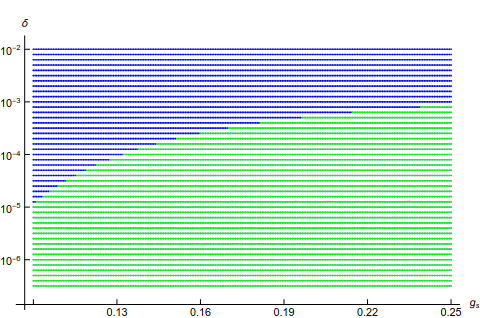}
        \caption{$\xi=1.3$}
        \label{fig:no_linear_inf_scan_naive-A4}
    \end{subfigure}
\caption{The green regions of the above parameter spaces (with $\delta\equiv W_X/W_0$) give a stable dS minimum for inflation, while in the blue regions the uplift becomes too large, leading to runaway potentials. This scan was done for the case without linear contribution setting $W_0=1$, $T_7=2\pi$ and $\alpha$ as in (\ref{ThetaDef}) with $\theta=0.994$, implying that in this setup we have both an inflationary and a late-time minimum.}
\label{fig:linear_inf_scan-Appendix}
\end{figure}

Let us now consider the case where the K\"ahler potential has no linear term in $X$, and study the effect of the corrections which have been ignored in Sec. \ref{subsec: no linear}. In this case the total scalar potential looks like:
\begin{equation}
V=\mathcal{A}\,\frac{W_X^2}{\sigma^2}+\mathcal{C}\,\frac{W_0^2}{\sigma^4}\,,
\end{equation}
where, as done in App. \ref{sec:Appendix Linear}, the coefficients $\mathcal{A}$ and $\mathcal{C}$ can be expanded order by order in perturbation theory, leading to a potential that can again be organised as the sum of LO, NLO and NNLO contributions. Fig. \ref{fig:linear_inf_scan-Appendix} shows the regions of the underlying $\{\xi, g_s,W_X \}$ parameter space which give rise to a stable dS minimum suitable for inflation, including all NNLO corrections. The explicit inclusion of known subdominant perturbative corrections merely requires a smaller value of $W_X$ but does not alter qualitatively the region of parameter space where we find an inflationary dS vacuum that is compatible with the late-time minimum.

\section{RG-induced moduli stabilisation and inflation}
\label{sec:RG Parameter Space}

In this appendix we review the RG-inspired moduli stabilisation scenario proposed in \cite{Burgess:2022nbx} with the goal in mind of deepening the analysis of brane-antibrane inflation. Let us consider a string-inspired supersymmetric 4D EFT where the superpotential $W$ does not depend on the K\"ahler modulus $T$, i.e. $W=W_0$, and the K\"ahler potential can be expanded as:
\begin{equation}
e^{-K/3}=\tau-k+ \ldots
\end{equation}
The key for this stabilisation mechanism is to assume that $k$ acquires a logarithmic dependence on the volume modulus through the running of a dimensionless coupling $\alpha_g$ (identifiable with the string coupling in the UV embedding). In fact, $k$ can be expanded perturbatively as:
\begin{equation}
k= k_0+k_1 \alpha_g +\frac{k_2}{2}\alpha_g^2+...
\label{kExpansion}
\end{equation}
and $\alpha_g$, being a coupling constant, is expected to receive quantum corrections which determine a renormalisation group running of the form (at first order):
\begin{equation}
\alpha_g(\tau)=\frac{\alpha_{g,0}}{1-b_1\alpha_{g,0}\ln\tau}\,,
\label{RGrunning}
\end{equation}
for some constant $b_1<0$ and $\alpha_{g,0}\equiv \alpha_g(\tau=1)$. The expression (\ref{RGrunning}) is justified by the fact that the running of $\alpha_g$ has a general logarithmic dependence on mass ratios which are functions of the volume modulus $\tau$. Combining (\ref{kExpansion}) with (\ref{RGrunning}), clearly induces a dependence of $k$ on $\ln\tau$.

\begin{figure}[t]
\centering
\includegraphics[scale=0.75]{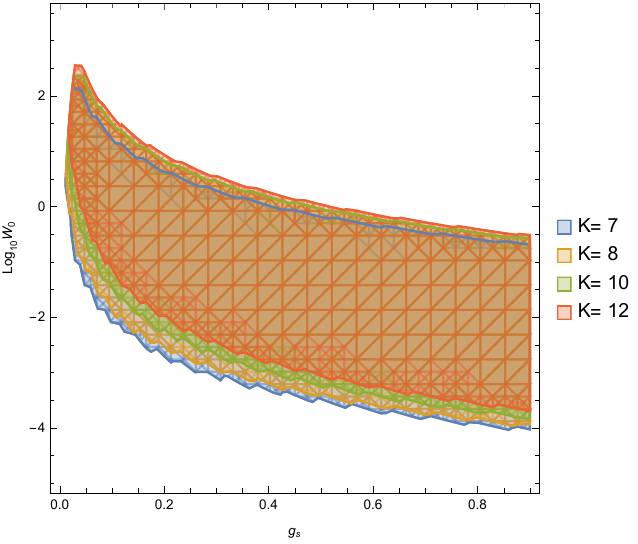}
\caption{Parameter space of the RG-induced inflation scenario consistent with the EFT constraints. As an estimate, for $K = 7$, $g_s= 0.1$ and $W_0 \simeq 0.1$, one has $M = 28$ and $\mathcal{V} \simeq 13000$.}
\label{fig:Paraspace}
\end{figure}

On the other hand, to incorporate brane-antibrane inflation, one can use the language of non-linear supersymmetry with a Goldstone fermion that encodes supersymmetry breaking through the chiral superfield $X$, constrained to satisfy $X^2=0$. In the same spirit, the inflaton $r$ is given by the scalar part of the chiral superfield $\Phi$ which satisfies $\overline{D}(X\overline{\Phi})=0$. This ensures that the superpotential and K\"ahler potential can be written as:
\begin{equation}
W=W_0(\Phi)+X W_X(\Phi, \overline{\Phi})\,,
\end{equation}
and:
\begin{equation}
\sigma\equiv e^{-K/3}=\tau-\hat{k}+... \,,  
\end{equation}
where $\hat{k}$ should reduce to $k$ in (\ref{kExpansion}) for $X=\Phi=0$, and can be expanded in powers of $X$ as:
\begin{eqnarray}
\hat{k}&=&\kappa(\Phi, \overline{\Phi}, \ln \tau)+(X+\overline{X})\kappa_X(\Phi, \overline{\Phi}, \ln \tau) \nonumber \\
&+&X\overline{X}\kappa_{X\overline{X}} (\Phi, \overline{\Phi}, \ln \tau)\,.
\end{eqnarray} 
Each function $\kappa$, $\kappa_X$ and $\kappa_{X\overline{X}}$ is expected to acquire a dependence on $\ln\tau$ from a perturbative series as (\ref{kExpansion}) combined with (\ref{RGrunning}) and with coefficients that depend on $\Phi$ and $\overline{\Phi}$. In particular, as pointed out in Sec. \ref{subsec:linear}, the absence of a linear dependence of $\hat{k}$ on $X$ at tree-level implies that $k_0=0$ for $\kappa_X$. The resulting scalar potential is:
\begin{equation}
V=\frac{1}{\sigma^2}\left[\frac13\kappa^{\overline{X}X}\left(W_X-3\kappa_{X \overline{T}} W_0\right)^2-3 \kappa_{T \overline{T}}W_0^2\right].
\label{RGPot}
\end{equation}
Each different contribution is expected to scale as follows: 
\begin{equation}
\kappa^{\overline{X}X}\sim\mathcal{O}(1)\quad \kappa_{X\overline{T}}\sim\mathcal{O}\left(\frac{\alpha_{g,0}^2}
{\sigma}\right) \quad
\kappa_{T \overline{T}} \sim\mathcal{O}\left(\frac{k_1\alpha_{g,0}^2}
{\sigma^2}\right),
\label{Scaling}
\end{equation}
where $\alpha_{g,0}\ll 1$ is the parameter controlling the perturbative expansion in (\ref{RGrunning}). As explained in Sec. \ref{subsec:linear}, the potential (\ref{RGPot}) can admit a stable dS minimum if the perfect square yields a leading order Minkowski minimum that is uplifted to dS by the correction proportional to $\kappa_{T \overline{T}}$. As can be seen from (\ref{Scaling}), this can indeed be the case only if the coefficient $k_1$ is negative and tuned to small values, $|k_1|\sim\mathcal{O}(\alpha_{g,0}^2)\ll 1$, so that $|\kappa_{T\overline{T}}|\lesssim \kappa^2_{X\overline{T}}$. Assuming that this tuning can indeed be performed, a dS minimum can then be found in the $\sigma$ direction at $\sigma_{\rm min}\sim  \alpha_{g,0}^2 W_0/W_X$ where $W_X\sim \kappa_{X\overline{T}} W_0$. For the inflationary period, the dominant term in the scalar potential is thus:
\begin{equation}
V_{\rm inf} \simeq \frac{|\kappa_{T\overline{T}}|\, W_0^2}{\sigma_{\rm min}^2}\sim \frac{W_X^2}{\sigma_{\rm min}^2}\,,
\label{eq:Vinf-parameterised}
\end{equation}
which can be identified with the brane-antibrane inflationary potential (\ref{InflPot}) if $W_X$ scales as (for $\mathcal{V}^{4/3}\simeq \sigma^2$):
\begin{equation}
W_X^2 \simeq \mathcal{C}_0 \equiv \frac{e^{-4 \rho}}{4 \pi}\,,
\end{equation}
where however now the volume mode is stabilised in a dS vacuum. As argued in \cite{Burgess:2022nbx}, the observational constraints on the amplitude of density perturbations and spectral index determine the inflation scale to be of order $V_{\rm inf}\simeq 10^{-17}$ (in Planck units). By using this constraint in (\ref{eq:Vinf-parameterised}), one obtains:
\begin{equation}
W_0 \simeq \frac{10^{17/2}}{4 \pi}\,\frac{e^{-4 \rho}}{g_s^2}\,,
\end{equation}
where we have identified for definiteness $\alpha_{g,0}$ with the string coupling, i.e. $\alpha_{g,0}=g_s\ll 1$. Moreover, since $\sigma \simeq \mathcal{V}^{2/3} \sim g_s^2 W_0/W_X$, we can rewrite the volume as:
\begin{equation}
\mathcal{V} \simeq \frac{10^{12.75}}{(4\pi)^{3/4}}\,e^{-3 \rho}\,.
\end{equation}
We can now derive the allowed regions of the underlying parameter space by requiring to have a consistent EFT which satisfies the theoretical conditions discussed in Sec. \ref{sec:Parameter-Space-Everything}. We plot in Fig. \ref{fig:Paraspace} the allowed regions of the $(W_0,g_s)$ parameter space for different values of the flux integer $K$.

\bibliographystyle{utphys}
\bibliography{BraneAntibraneInflV2}

\providecommand{\href}[2]{#2}\begingroup\raggedright\begin{thebibliography}{100}

\bibitem{Planck:2018jri}
{\bfseries Planck} Collaboration, Y.~Akrami {\em et~al.}, ``{Planck 2018 results. X. Constraints on inflation},'' \href{http://dx.doi.org/10.1051/0004-6361/201833887}{{\em Astron. Astrophys.} {\bfseries 641} (2020) A10}, \href{http://arxiv.org/abs/1807.06211}{{\ttfamily arXiv:1807.06211 [astro-ph.CO]}}.

\bibitem{Martin:2024qnn}
J.~Martin, C.~Ringeval, and V.~Vennin, ``{Cosmic Inflation at the Crossroads},'' \href{http://arxiv.org/abs/2404.10647}{{\ttfamily arXiv:2404.10647 [astro-ph.CO]}}.

\bibitem{Burgess:2014tja}
C.~P. Burgess, M.~Cicoli, F.~Quevedo, and M.~Williams, ``{Inflating with Large Effective Fields},'' \href{http://dx.doi.org/10.1088/1475-7516/2014/11/045}{{\em JCAP} {\bfseries 11} (2014) 045}, \href{http://arxiv.org/abs/1404.6236}{{\ttfamily arXiv:1404.6236 [hep-th]}}.

\bibitem{Cicoli:2023opf}
M.~Cicoli, J.~P. Conlon, A.~Maharana, S.~Parameswaran, F.~Quevedo, and I.~Zavala, ``{String cosmology: From the early universe to today},'' \href{http://dx.doi.org/10.1016/j.physrep.2024.01.002}{{\em Phys. Rept.} {\bfseries 1059} (2024) 1--155}, \href{http://arxiv.org/abs/2303.04819}{{\ttfamily arXiv:2303.04819 [hep-th]}}.

\bibitem{Bansal:2024uzr}
S.~Bansal, L.~Brunelli, M.~Cicoli, A.~Hebecker, and R.~Kuespert, ``{Loop Blow-up Inflation},'' \href{http://arxiv.org/abs/2403.04831}{{\ttfamily arXiv:2403.04831 [hep-th]}}.

\bibitem{Cicoli:2008gp}
M.~Cicoli, C.~P. Burgess, and F.~Quevedo, ``{Fibre Inflation: Observable Gravity Waves from IIB String Compactifications},'' \href{http://dx.doi.org/10.1088/1475-7516/2009/03/013}{{\em JCAP} {\bfseries 03} (2009) 013}, \href{http://arxiv.org/abs/0808.0691}{{\ttfamily arXiv:0808.0691 [hep-th]}}.

\bibitem{Broy:2015zba}
B.~J. Broy, D.~Ciupke, F.~G. Pedro, and A.~Westphal, ``{Starobinsky-Type Inflation from $\alpha'$-Corrections},'' \href{http://dx.doi.org/10.1088/1475-7516/2016/01/001}{{\em JCAP} {\bfseries 01} (2016) 001}, \href{http://arxiv.org/abs/1509.00024}{{\ttfamily arXiv:1509.00024 [hep-th]}}.

\bibitem{Cicoli:2016chb}
M.~Cicoli, D.~Ciupke, S.~de~Alwis, and F.~Muia, ``{$\alpha'$ Inflation: moduli stabilisation and observable tensors from higher derivatives},'' \href{http://dx.doi.org/10.1007/JHEP09(2016)026}{{\em JHEP} {\bfseries 09} (2016) 026}, \href{http://arxiv.org/abs/1607.01395}{{\ttfamily arXiv:1607.01395 [hep-th]}}.

\bibitem{Cicoli:2016xae}
M.~Cicoli, F.~Muia, and P.~Shukla, ``{Global Embedding of Fibre Inflation Models},'' \href{http://dx.doi.org/10.1007/JHEP11(2016)182}{{\em JHEP} {\bfseries 11} (2016) 182}, \href{http://arxiv.org/abs/1611.04612}{{\ttfamily arXiv:1611.04612 [hep-th]}}.

\bibitem{Cicoli:2017axo}
M.~Cicoli, D.~Ciupke, V.~A. Diaz, V.~Guidetti, F.~Muia, and P.~Shukla, ``{Chiral Global Embedding of Fibre Inflation Models},'' \href{http://dx.doi.org/10.1007/JHEP11(2017)207}{{\em JHEP} {\bfseries 11} (2017) 207}, \href{http://arxiv.org/abs/1709.01518}{{\ttfamily arXiv:1709.01518 [hep-th]}}.

\bibitem{Cicoli:2020bao}
M.~Cicoli and E.~Di~Valentino, ``{Fitting string inflation to real cosmological data: The fiber inflation case},'' \href{http://dx.doi.org/10.1103/PhysRevD.102.043521}{{\em Phys. Rev. D} {\bfseries 102} no.~4, (2020) 043521}, \href{http://arxiv.org/abs/2004.01210}{{\ttfamily arXiv:2004.01210 [astro-ph.CO]}}.

\bibitem{Burgess:2016owb}
C.~P. Burgess, M.~Cicoli, S.~de~Alwis, and F.~Quevedo, ``{Robust Inflation from Fibrous Strings},'' \href{http://dx.doi.org/10.1088/1475-7516/2016/05/032}{{\em JCAP} {\bfseries 05} (2016) 032}, \href{http://arxiv.org/abs/1603.06789}{{\ttfamily arXiv:1603.06789 [hep-th]}}.

\bibitem{Brinkmann:2023eph}
M.~Brinkmann, M.~Cicoli, and P.~Zito, ``{Starobinsky inflation from string theory?},'' \href{http://dx.doi.org/10.1007/JHEP09(2023)038}{{\em JHEP} {\bfseries 09} (2023) 038}, \href{http://arxiv.org/abs/2305.05703}{{\ttfamily arXiv:2305.05703 [hep-th]}}.

\bibitem{Lust:2023zql}
D.~Lust, J.~Masias, B.~Muntz, and M.~Scalisi, ``{Starobinsky Inflation in the Swampland},'' \href{http://arxiv.org/abs/2312.13210}{{\ttfamily arXiv:2312.13210 [hep-th]}}.

\bibitem{Dvali:1998pa}
G.~R. Dvali and S.~H.~H. Tye, ``{Brane inflation},'' \href{http://dx.doi.org/10.1016/S0370-2693(99)00132-X}{{\em Phys. Lett. B} {\bfseries 450} (1999) 72--82}, \href{http://arxiv.org/abs/hep-ph/9812483}{{\ttfamily arXiv:hep-ph/9812483}}.

\bibitem{Burgess:2001fx}
C.~P. Burgess, M.~Majumdar, D.~Nolte, F.~Quevedo, G.~Rajesh, and R.-J. Zhang, ``{The Inflationary brane anti-brane universe},'' \href{http://dx.doi.org/10.1088/1126-6708/2001/07/047}{{\em JHEP} {\bfseries 07} (2001) 047}, \href{http://arxiv.org/abs/hep-th/0105204}{{\ttfamily arXiv:hep-th/0105204}}.

\bibitem{Dvali:2001fw}
G.~R. Dvali, Q.~Shafi, and S.~Solganik, ``{D-brane inflation},'' in {\em {4th European Meeting From the Planck Scale to the Electroweak Scale}}.
\newblock 5, 2001.
\newblock \href{http://arxiv.org/abs/hep-th/0105203}{{\ttfamily arXiv:hep-th/0105203}}.

\bibitem{Kachru:2003sx}
S.~Kachru, R.~Kallosh, A.~D. Linde, J.~M. Maldacena, L.~P. McAllister, and S.~P. Trivedi, ``{Towards inflation in string theory},'' \href{http://dx.doi.org/10.1088/1475-7516/2003/10/013}{{\em JCAP} {\bfseries 10} (2003) 013}, \href{http://arxiv.org/abs/hep-th/0308055}{{\ttfamily arXiv:hep-th/0308055}}.

\bibitem{Baumann:2007ah}
D.~Baumann, A.~Dymarsky, I.~R. Klebanov, and L.~McAllister, ``{Towards an Explicit Model of D-brane Inflation},'' \href{http://dx.doi.org/10.1088/1475-7516/2008/01/024}{{\em JCAP} {\bfseries 01} (2008) 024}, \href{http://arxiv.org/abs/0706.0360}{{\ttfamily arXiv:0706.0360 [hep-th]}}.

\bibitem{Burgess:2022nbx}
C.~P. Burgess and F.~Quevedo, ``{RG-induced modulus stabilization: perturbative de Sitter vacua and improved D3-$ \overline{\mathrm{D}3} $ inflation},'' \href{http://dx.doi.org/10.1007/JHEP06(2022)167}{{\em JHEP} {\bfseries 06} (2022) 167}, \href{http://arxiv.org/abs/2202.05344}{{\ttfamily arXiv:2202.05344 [hep-th]}}.

\bibitem{Weissenbacher:2019mef}
M.~Weissenbacher, ``{F-theory vacua and $\alpha'$-corrections},'' \href{http://dx.doi.org/10.1007/JHEP04(2020)032}{{\em JHEP} {\bfseries 04} (2020) 032}, \href{http://arxiv.org/abs/1901.04758}{{\ttfamily arXiv:1901.04758 [hep-th]}}.

\bibitem{Weissenbacher:2020cyf}
M.~Weissenbacher, ``{On $\alpha'$-effects from $D$-branes in $4d \; \mathcal{N} = 1$},'' \href{http://dx.doi.org/10.1007/JHEP11(2020)076}{{\em JHEP} {\bfseries 11} (2020) 076}, \href{http://arxiv.org/abs/2006.15552}{{\ttfamily arXiv:2006.15552 [hep-th]}}.

\bibitem{Klaewer:2020lfg}
D.~Klaewer, S.-J. Lee, T.~Weigand, and M.~Wiesner, ``{Quantum corrections in 4d $N$ = 1 infinite distance limits and the weak gravity conjecture},'' \href{http://dx.doi.org/10.1007/JHEP03(2021)252}{{\em JHEP} {\bfseries 03} (2021) 252}, \href{http://arxiv.org/abs/2011.00024}{{\ttfamily arXiv:2011.00024 [hep-th]}}.

\bibitem{Becker:2002nn}
K.~Becker, M.~Becker, M.~Haack, and J.~Louis, ``{Supersymmetry breaking and alpha-prime corrections to flux induced potentials},'' \href{http://dx.doi.org/10.1088/1126-6708/2002/06/060}{{\em JHEP} {\bfseries 06} (2002) 060}, \href{http://arxiv.org/abs/hep-th/0204254}{{\ttfamily arXiv:hep-th/0204254}}.

\bibitem{Bonetti:2016dqh}
F.~Bonetti and M.~Weissenbacher, ``{The Euler characteristic correction to the K\"ahler potential \textemdash{} revisited},'' \href{http://dx.doi.org/10.1007/JHEP01(2017)003}{{\em JHEP} {\bfseries 01} (2017) 003}, \href{http://arxiv.org/abs/1608.01300}{{\ttfamily arXiv:1608.01300 [hep-th]}}.

\bibitem{Antoniadis:2018hqy}
I.~Antoniadis, Y.~Chen, and G.~K. Leontaris, ``{Perturbative moduli stabilisation in type IIB/F-theory framework},'' \href{http://dx.doi.org/10.1140/epjc/s10052-018-6248-4}{{\em Eur. Phys. J. C} {\bfseries 78} no.~9, (2018) 766}, \href{http://arxiv.org/abs/1803.08941}{{\ttfamily arXiv:1803.08941 [hep-th]}}.

\bibitem{Antoniadis:2019rkh}
I.~Antoniadis, Y.~Chen, and G.~K. Leontaris, ``{Logarithmic loop corrections, moduli stabilisation and de Sitter vacua in string theory},'' \href{http://dx.doi.org/10.1007/JHEP01(2020)149}{{\em JHEP} {\bfseries 01} (2020) 149}, \href{http://arxiv.org/abs/1909.10525}{{\ttfamily arXiv:1909.10525 [hep-th]}}.

\bibitem{Conlon:2022pnx}
J.~P. Conlon and F.~Revello, ``{Catch-me-if-you-can: the overshoot problem and the weak/inflation hierarchy},'' \href{http://dx.doi.org/10.1007/JHEP11(2022)155}{{\em JHEP} {\bfseries 11} (2022) 155}, \href{http://arxiv.org/abs/2207.00567}{{\ttfamily arXiv:2207.00567 [hep-th]}}.

\bibitem{Apers:2022cyl}
F.~Apers, J.~P. Conlon, M.~Mosny, and F.~Revello, ``{Kination, meet Kasner: on the asymptotic cosmology of string compactifications},'' \href{http://dx.doi.org/10.1007/JHEP08(2023)156}{{\em JHEP} {\bfseries 08} (2023) 156}, \href{http://arxiv.org/abs/2212.10293}{{\ttfamily arXiv:2212.10293 [hep-th]}}.

\bibitem{Apers:2024ffe}
F.~Apers, J.~P. Conlon, E.~J. Copeland, M.~Mosny, and F.~Revello, ``{String Theory and the First Half of the Universe},'' \href{http://arxiv.org/abs/2401.04064}{{\ttfamily arXiv:2401.04064 [hep-th]}}.

\bibitem{Quevedo:2002xw}
F.~Quevedo, ``{Lectures on string/brane cosmology},'' \href{http://dx.doi.org/10.1088/0264-9381/19/22/304}{{\em Class. Quant. Grav.} {\bfseries 19} (2002) 5721--5779}, \href{http://arxiv.org/abs/hep-th/0210292}{{\ttfamily arXiv:hep-th/0210292}}.

\bibitem{Baumann:2014nda}
D.~Baumann and L.~McAllister, \href{http://dx.doi.org/10.1017/CBO9781316105733}{{\em {Inflation and String Theory}}}.
\newblock Cambridge Monographs on Mathematical Physics. Cambridge University Press, 5, 2015.
\newblock \href{http://arxiv.org/abs/1404.2601}{{\ttfamily arXiv:1404.2601 [hep-th]}}.

\bibitem{Tye:2023rwa}
H.~S.~H. Tye, ``{The D3-$ \overline{D}3 $-brane inflation model revisited},'' \href{http://dx.doi.org/10.1007/JHEP01(2024)171}{{\em JHEP} {\bfseries 01} (2024) 171}, \href{http://arxiv.org/abs/2311.07810}{{\ttfamily arXiv:2311.07810 [hep-th]}}.

\bibitem{Cicoli:2011yy}
M.~Cicoli, C.~P. Burgess, and F.~Quevedo, ``{Anisotropic Modulus Stabilisation: Strings at LHC Scales with Micron-sized Extra Dimensions},'' \href{http://dx.doi.org/10.1007/JHEP10(2011)119}{{\em JHEP} {\bfseries 10} (2011) 119}, \href{http://arxiv.org/abs/1105.2107}{{\ttfamily arXiv:1105.2107 [hep-th]}}.

\bibitem{Aparicio:2015psl}
L.~Aparicio, F.~Quevedo, and R.~Valandro, ``{Moduli Stabilisation with Nilpotent Goldstino: Vacuum Structure and SUSY Breaking},'' \href{http://dx.doi.org/10.1007/JHEP03(2016)036}{{\em JHEP} {\bfseries 03} (2016) 036}, \href{http://arxiv.org/abs/1511.08105}{{\ttfamily arXiv:1511.08105 [hep-th]}}.

\bibitem{Liu:2022bfg}
J.~T. Liu, R.~Minasian, R.~Savelli, and A.~Schachner, ``{Type IIB at eight derivatives: insights from Superstrings, Superfields and Superparticles},'' \href{http://dx.doi.org/10.1007/JHEP08(2022)267}{{\em JHEP} {\bfseries 08} (2022) 267}, \href{http://arxiv.org/abs/2205.11530}{{\ttfamily arXiv:2205.11530 [hep-th]}}.

\bibitem{Wulff:2021fhr}
L.~Wulff, ``{Completing R$^{4}$ using O(d, d)},'' \href{http://dx.doi.org/10.1007/JHEP08(2022)187}{{\em JHEP} {\bfseries 08} (2022) 187}, \href{http://arxiv.org/abs/2111.00018}{{\ttfamily arXiv:2111.00018 [hep-th]}}.

\bibitem{Wulff:2024mgu}
L.~Wulff, ``{Tree-level R$^{4}$ correction from O(d, d): NS-NS five-point terms},'' \href{http://dx.doi.org/10.1007/JHEP09(2024)078}{{\em JHEP} {\bfseries 09} (2024) 078}, \href{http://arxiv.org/abs/2406.15240}{{\ttfamily arXiv:2406.15240 [hep-th]}}.

\bibitem{Cicoli:2021rub}
M.~Cicoli, F.~Quevedo, R.~Savelli, A.~Schachner, and R.~Valandro, ``{Systematics of the \ensuremath{\alpha}' expansion in F-theory},'' \href{http://dx.doi.org/10.1007/JHEP08(2021)099}{{\em JHEP} {\bfseries 08} (2021) 099}, \href{http://arxiv.org/abs/2106.04592}{{\ttfamily arXiv:2106.04592 [hep-th]}}.

\bibitem{Burgess:2020qsc}
C.~P. Burgess, M.~Cicoli, D.~Ciupke, S.~Krippendorf, and F.~Quevedo, ``{UV Shadows in EFTs: Accidental Symmetries, Robustness and No-Scale Supergravity},'' \href{http://dx.doi.org/10.1002/prop.202000076}{{\em Fortsch. Phys.} {\bfseries 68} no.~10, (2020) 2000076}, \href{http://arxiv.org/abs/2006.06694}{{\ttfamily arXiv:2006.06694 [hep-th]}}.

\bibitem{Grimm:2013gma}
T.~W. Grimm, R.~Savelli, and M.~Weissenbacher, ``{On \textbackslash{}alpha' corrections in N=1 F-theory compactifications},'' \href{http://dx.doi.org/10.1016/j.physletb.2013.07.024}{{\em Phys. Lett. B} {\bfseries 725} (2013) 431--436}, \href{http://arxiv.org/abs/1303.3317}{{\ttfamily arXiv:1303.3317 [hep-th]}}.

\bibitem{Conlon:2009xf}
J.~P. Conlon, ``{Gauge Threshold Corrections for Local String Models},'' \href{http://dx.doi.org/10.1088/1126-6708/2009/04/059}{{\em JHEP} {\bfseries 04} (2009) 059}, \href{http://arxiv.org/abs/0901.4350}{{\ttfamily arXiv:0901.4350 [hep-th]}}.

\bibitem{Conlon:2009qa}
J.~P. Conlon and E.~Palti, ``{On Gauge Threshold Corrections for Local IIB/F-theory GUTs},'' \href{http://dx.doi.org/10.1103/PhysRevD.80.106004}{{\em Phys. Rev. D} {\bfseries 80} (2009) 106004}, \href{http://arxiv.org/abs/0907.1362}{{\ttfamily arXiv:0907.1362 [hep-th]}}.

\bibitem{Conlon:2009kt}
J.~P. Conlon and E.~Palti, ``{Gauge Threshold Corrections for Local Orientifolds},'' \href{http://dx.doi.org/10.1088/1126-6708/2009/09/019}{{\em JHEP} {\bfseries 09} (2009) 019}, \href{http://arxiv.org/abs/0906.1920}{{\ttfamily arXiv:0906.1920 [hep-th]}}.

\bibitem{Berg:2005ja}
M.~Berg, M.~Haack, and B.~Kors, ``{String loop corrections to Kahler potentials in orientifolds},'' \href{http://dx.doi.org/10.1088/1126-6708/2005/11/030}{{\em JHEP} {\bfseries 11} (2005) 030}, \href{http://arxiv.org/abs/hep-th/0508043}{{\ttfamily arXiv:hep-th/0508043}}.

\bibitem{Berg:2007wt}
M.~Berg, M.~Haack, and E.~Pajer, ``{Jumping Through Loops: On Soft Terms from Large Volume Compactifications},'' \href{http://dx.doi.org/10.1088/1126-6708/2007/09/031}{{\em JHEP} {\bfseries 09} (2007) 031}, \href{http://arxiv.org/abs/0704.0737}{{\ttfamily arXiv:0704.0737 [hep-th]}}.

\bibitem{Cicoli:2007xp}
M.~Cicoli, J.~P. Conlon, and F.~Quevedo, ``{Systematics of String Loop Corrections in Type IIB Calabi-Yau Flux Compactifications},'' \href{http://dx.doi.org/10.1088/1126-6708/2008/01/052}{{\em JHEP} {\bfseries 01} (2008) 052}, \href{http://arxiv.org/abs/0708.1873}{{\ttfamily arXiv:0708.1873 [hep-th]}}.

\bibitem{Minasian:2015bxa}
R.~Minasian, T.~G. Pugh, and R.~Savelli, ``{F-theory at order $\alpha'^3$},'' \href{http://dx.doi.org/10.1007/JHEP10(2015)050}{{\em JHEP} {\bfseries 10} (2015) 050}, \href{http://arxiv.org/abs/1506.06756}{{\ttfamily arXiv:1506.06756 [hep-th]}}.

\bibitem{Antoniadis:1997eg}
I.~Antoniadis, S.~Ferrara, R.~Minasian, and K.~S. Narain, ``{R**4 couplings in M and type II theories on Calabi-Yau spaces},'' \href{http://dx.doi.org/10.1016/S0550-3213(97)00572-5}{{\em Nucl. Phys. B} {\bfseries 507} (1997) 571--588}, \href{http://arxiv.org/abs/hep-th/9707013}{{\ttfamily arXiv:hep-th/9707013}}.

\bibitem{Leontaris:2022rzj}
G.~K. Leontaris and P.~Shukla, ``{Stabilising all K\"ahler moduli in perturbative LVS},'' \href{http://dx.doi.org/10.1007/JHEP07(2022)047}{{\em JHEP} {\bfseries 07} (2022) 047}, \href{http://arxiv.org/abs/2203.03362}{{\ttfamily arXiv:2203.03362 [hep-th]}}.

\bibitem{Berg:2014ama}
M.~Berg, M.~Haack, J.~U. Kang, and S.~Sj\"ors, ``{Towards the one-loop K\"ahler metric of Calabi-Yau orientifolds},'' \href{http://dx.doi.org/10.1007/JHEP12(2014)077}{{\em JHEP} {\bfseries 12} (2014) 077}, \href{http://arxiv.org/abs/1407.0027}{{\ttfamily arXiv:1407.0027 [hep-th]}}.

\bibitem{Haack:2015pbv}
M.~Haack and J.~U. Kang, ``{One-loop Einstein-Hilbert term in minimally supersymmetric type IIB orientifolds},'' \href{http://dx.doi.org/10.1007/JHEP02(2016)160}{{\em JHEP} {\bfseries 02} (2016) 160}, \href{http://arxiv.org/abs/1511.03957}{{\ttfamily arXiv:1511.03957 [hep-th]}}.

\bibitem{Ciupke:2015msa}
D.~Ciupke, J.~Louis, and A.~Westphal, ``{Higher-Derivative Supergravity and Moduli Stabilization},'' \href{http://dx.doi.org/10.1007/JHEP10(2015)094}{{\em JHEP} {\bfseries 10} (2015) 094}, \href{http://arxiv.org/abs/1505.03092}{{\ttfamily arXiv:1505.03092 [hep-th]}}.

\bibitem{Grimm:2017okk}
T.~W. Grimm, K.~Mayer, and M.~Weissenbacher, ``{Higher derivatives in Type II and M-theory on Calabi-Yau threefolds},'' \href{http://dx.doi.org/10.1007/JHEP02(2018)127}{{\em JHEP} {\bfseries 02} (2018) 127}, \href{http://arxiv.org/abs/1702.08404}{{\ttfamily arXiv:1702.08404 [hep-th]}}.

\bibitem{Cicoli:2023njy}
M.~Cicoli, M.~Licheri, P.~Piantadosi, F.~Quevedo, and P.~Shukla, ``{Higher derivative corrections to string inflation},'' \href{http://dx.doi.org/10.1007/JHEP02(2024)115}{{\em JHEP} {\bfseries 02} (2024) 115}, \href{http://arxiv.org/abs/2309.11697}{{\ttfamily arXiv:2309.11697 [hep-th]}}.

\bibitem{deAlwis:2012vp}
S.~P. de~Alwis, ``{Constraints on LVS Compactifications of IIB String Theory},'' \href{http://dx.doi.org/10.1007/JHEP05(2012)026}{{\em JHEP} {\bfseries 05} (2012) 026}, \href{http://arxiv.org/abs/1202.1546}{{\ttfamily arXiv:1202.1546 [hep-th]}}.

\bibitem{Cicoli:2013swa}
M.~Cicoli, J.~P. Conlon, A.~Maharana, and F.~Quevedo, ``{A Note on the Magnitude of the Flux Superpotential},'' \href{http://dx.doi.org/10.1007/JHEP01(2014)027}{{\em JHEP} {\bfseries 01} (2014) 027}, \href{http://arxiv.org/abs/1310.6694}{{\ttfamily arXiv:1310.6694 [hep-th]}}.

\bibitem{Gao:2022uop}
X.~Gao, A.~Hebecker, S.~Schreyer, and G.~Venken, ``{Loops, local corrections and warping in the LVS and other type IIB models},'' \href{http://dx.doi.org/10.1007/JHEP09(2022)091}{{\em JHEP} {\bfseries 09} (2022) 091}, \href{http://arxiv.org/abs/2204.06009}{{\ttfamily arXiv:2204.06009 [hep-th]}}.

\bibitem{vonGersdorff:2005bf}
G.~von Gersdorff and A.~Hebecker, ``{Kahler corrections for the volume modulus of flux compactifications},'' \href{http://dx.doi.org/10.1016/j.physletb.2005.08.024}{{\em Phys. Lett. B} {\bfseries 624} (2005) 270--274}, \href{http://arxiv.org/abs/hep-th/0507131}{{\ttfamily arXiv:hep-th/0507131}}.

\bibitem{Berg:2005yu}
M.~Berg, M.~Haack, and B.~Kors, ``{On volume stabilization by quantum corrections},'' \href{http://dx.doi.org/10.1103/PhysRevLett.96.021601}{{\em Phys. Rev. Lett.} {\bfseries 96} (2006) 021601}, \href{http://arxiv.org/abs/hep-th/0508171}{{\ttfamily arXiv:hep-th/0508171}}.

\bibitem{Balasubramanian:2005zx}
V.~Balasubramanian, P.~Berglund, J.~P. Conlon, and F.~Quevedo, ``{Systematics of moduli stabilisation in Calabi-Yau flux compactifications},'' \href{http://dx.doi.org/10.1088/1126-6708/2005/03/007}{{\em JHEP} {\bfseries 03} (2005) 007}, \href{http://arxiv.org/abs/hep-th/0502058}{{\ttfamily arXiv:hep-th/0502058}}.

\bibitem{Cicoli:2015ylx}
M.~Cicoli, F.~Quevedo, and R.~Valandro, ``{De Sitter from T-branes},'' \href{http://dx.doi.org/10.1007/JHEP03(2016)141}{{\em JHEP} {\bfseries 03} (2016) 141}, \href{http://arxiv.org/abs/1512.04558}{{\ttfamily arXiv:1512.04558 [hep-th]}}.

\bibitem{Gallego:2017dvd}
D.~Gallego, M.~C.~D. Marsh, B.~Vercnocke, and T.~Wrase, ``{A New Class of de Sitter Vacua in Type IIB Large Volume Compactifications},'' \href{http://dx.doi.org/10.1007/JHEP10(2017)193}{{\em JHEP} {\bfseries 10} (2017) 193}, \href{http://arxiv.org/abs/1707.01095}{{\ttfamily arXiv:1707.01095 [hep-th]}}.

\bibitem{Burgess:2003ic}
C.~P. Burgess, R.~Kallosh, and F.~Quevedo, ``{De Sitter string vacua from supersymmetric D terms},'' \href{http://dx.doi.org/10.1088/1126-6708/2003/10/056}{{\em JHEP} {\bfseries 10} (2003) 056}, \href{http://arxiv.org/abs/hep-th/0309187}{{\ttfamily arXiv:hep-th/0309187}}.

\bibitem{Cicoli:2012fh}
M.~Cicoli, A.~Maharana, F.~Quevedo, and C.~P. Burgess, ``{De Sitter String Vacua from Dilaton-dependent Non-perturbative Effects},'' \href{http://dx.doi.org/10.1007/JHEP06(2012)011}{{\em JHEP} {\bfseries 06} (2012) 011}, \href{http://arxiv.org/abs/1203.1750}{{\ttfamily arXiv:1203.1750 [hep-th]}}.

\bibitem{Kreuzer:2000xy}
M.~Kreuzer and H.~Skarke, ``{Complete classification of reflexive polyhedra in four-dimensions},'' \href{http://dx.doi.org/10.4310/ATMP.2000.v4.n6.a2}{{\em Adv. Theor. Math. Phys.} {\bfseries 4} (2000) 1209--1230}, \href{http://arxiv.org/abs/hep-th/0002240}{{\ttfamily arXiv:hep-th/0002240}}.

\bibitem{Cicoli:2017shd}
M.~Cicoli, I.~n. Garc\`\i{}a-Etxebarria, C.~Mayrhofer, F.~Quevedo, P.~Shukla, and R.~Valandro, ``{Global Orientifolded Quivers with Inflation},'' \href{http://dx.doi.org/10.1007/JHEP11(2017)134}{{\em JHEP} {\bfseries 11} (2017) 134}, \href{http://arxiv.org/abs/1706.06128}{{\ttfamily arXiv:1706.06128 [hep-th]}}.

\bibitem{Cicoli:2021dhg}
M.~Cicoli, I.~n.~G. Etxebarria, F.~Quevedo, A.~Schachner, P.~Shukla, and R.~Valandro, ``{The Standard Model quiver in de Sitter string compactifications},'' \href{http://dx.doi.org/10.1007/JHEP08(2021)109}{{\em JHEP} {\bfseries 08} (2021) 109}, \href{http://arxiv.org/abs/2106.11964}{{\ttfamily arXiv:2106.11964 [hep-th]}}.

\bibitem{Antoniadis:2024hvw}
I.~Antoniadis, E.~Dudas, F.~Farakos, and A.~Sagnotti, {\em {Non-Linear Supergravity and Inflationary Cosmology}}.
\newblock 9, 2024.
\newblock \href{http://arxiv.org/abs/2409.14943}{{\ttfamily arXiv:2409.14943 [hep-th]}}.

\bibitem{Guleryuz:2022hiv}
O.~Guleryuz, ``{(Super)universal attractors and the de~Sitter vacua in string landscape},'' \href{http://dx.doi.org/10.1088/1475-7516/2023/05/039}{{\em JCAP} {\bfseries 05} (2023) 039}, \href{http://arxiv.org/abs/2207.10634}{{\ttfamily arXiv:2207.10634 [hep-th]}}.

\bibitem{Burgess:2021obw}
C.~P. Burgess, D.~Dineen, and F.~Quevedo, ``{Yoga Dark Energy: natural relaxation and other dark implications of a supersymmetric gravity sector},'' \href{http://dx.doi.org/10.1088/1475-7516/2022/03/064}{{\em JCAP} {\bfseries 03} no.~03, (2022) 064}, \href{http://arxiv.org/abs/2111.07286}{{\ttfamily arXiv:2111.07286 [hep-th]}}.

\bibitem{DallAgata:2015zxp}
G.~Dall'Agata and F.~Farakos, ``{Constrained superfields in Supergravity},'' \href{http://dx.doi.org/10.1007/JHEP02(2016)101}{{\em JHEP} {\bfseries 02} (2016) 101}, \href{http://arxiv.org/abs/1512.02158}{{\ttfamily arXiv:1512.02158 [hep-th]}}.

\bibitem{McDonough:2016der}
E.~McDonough and M.~Scalisi, ``{Inflation from Nilpotent K\"ahler Corrections},'' \href{http://dx.doi.org/10.1088/1475-7516/2016/11/028}{{\em JCAP} {\bfseries 11} (2016) 028}, \href{http://arxiv.org/abs/1609.00364}{{\ttfamily arXiv:1609.00364 [hep-th]}}.

\bibitem{Kallosh:2017wnt}
R.~Kallosh, A.~Linde, D.~Roest, and Y.~Yamada, ``{$ \overline{D3} $ induced geometric inflation},'' \href{http://dx.doi.org/10.1007/JHEP07(2017)057}{{\em JHEP} {\bfseries 07} (2017) 057}, \href{http://arxiv.org/abs/1705.09247}{{\ttfamily arXiv:1705.09247 [hep-th]}}.

\bibitem{Gao:2022fdi}
X.~Gao, A.~Hebecker, S.~Schreyer, and G.~Venken, ``{The LVS parametric tadpole constraint},'' \href{http://dx.doi.org/10.1007/JHEP07(2022)056}{{\em JHEP} {\bfseries 07} (2022) 056}, \href{http://arxiv.org/abs/2202.04087}{{\ttfamily arXiv:2202.04087 [hep-th]}}.

\bibitem{Gao:2020xqh}
X.~Gao, A.~Hebecker, and D.~Junghans, ``{Control issues of KKLT},'' \href{http://dx.doi.org/10.1002/prop.202000089}{{\em Fortsch. Phys.} {\bfseries 68} (2020) 2000089}, \href{http://arxiv.org/abs/2009.03914}{{\ttfamily arXiv:2009.03914 [hep-th]}}.

\bibitem{Carta:2019rhx}
F.~Carta, J.~Moritz, and A.~Westphal, ``{Gaugino condensation and small uplifts in KKLT},'' \href{http://dx.doi.org/10.1007/JHEP08(2019)141}{{\em JHEP} {\bfseries 08} (2019) 141}, \href{http://arxiv.org/abs/1902.01412}{{\ttfamily arXiv:1902.01412 [hep-th]}}.

\bibitem{Junghans:2022exo}
D.~Junghans, ``{LVS de Sitter vacua are probably in the swampland},'' \href{http://dx.doi.org/10.1016/j.nuclphysb.2023.116179}{{\em Nucl. Phys. B} {\bfseries 990} (2023) 116179}, \href{http://arxiv.org/abs/2201.03572}{{\ttfamily arXiv:2201.03572 [hep-th]}}.

\bibitem{Junghans:2022kxg}
D.~Junghans, ``{Topological constraints in the LARGE-volume scenario},'' \href{http://dx.doi.org/10.1007/JHEP08(2022)226}{{\em JHEP} {\bfseries 08} (2022) 226}, \href{http://arxiv.org/abs/2205.02856}{{\ttfamily arXiv:2205.02856 [hep-th]}}.

\bibitem{Hebecker:2022zme}
A.~Hebecker, S.~Schreyer, and G.~Venken, ``{Curvature corrections to KPV: do we need deep throats?},'' \href{http://dx.doi.org/10.1007/JHEP10(2022)166}{{\em JHEP} {\bfseries 10} (2022) 166}, \href{http://arxiv.org/abs/2208.02826}{{\ttfamily arXiv:2208.02826 [hep-th]}}.

\bibitem{Schreyer:2022len}
S.~Schreyer and G.~Venken, ``{\ensuremath{\alpha}' corrections to KPV: an uplifting story},'' \href{http://dx.doi.org/10.1007/JHEP07(2023)235}{{\em JHEP} {\bfseries 07} (2023) 235}, \href{http://arxiv.org/abs/2212.07437}{{\ttfamily arXiv:2212.07437 [hep-th]}}.

\bibitem{Schreyer:2024pml}
S.~Schreyer, ``{Higher order corrections to KPV: The nonabelian brane stack perspective},'' \href{http://arxiv.org/abs/2402.13311}{{\ttfamily arXiv:2402.13311 [hep-th]}}.

\bibitem{Kachru:2002gs}
S.~Kachru, J.~Pearson, and H.~L. Verlinde, ``{Brane / flux annihilation and the string dual of a nonsupersymmetric field theory},'' \href{http://dx.doi.org/10.1088/1126-6708/2002/06/021}{{\em JHEP} {\bfseries 06} (2002) 021}, \href{http://arxiv.org/abs/hep-th/0112197}{{\ttfamily arXiv:hep-th/0112197}}.

\bibitem{Bena:2018fqc}
I.~Bena, E.~Dudas, M.~Gra\~na, and S.~L\"ust, ``{Uplifting Runaways},'' \href{http://dx.doi.org/10.1002/prop.201800100}{{\em Fortsch. Phys.} {\bfseries 67} no.~1-2, (2019) 1800100}, \href{http://arxiv.org/abs/1809.06861}{{\ttfamily arXiv:1809.06861 [hep-th]}}.

\bibitem{Lust:2022xoq}
S.~L\"ust and L.~Randall, ``{Effective Theory of Warped Compactifications and the Implications for KKLT},'' \href{http://dx.doi.org/10.1002/prop.202200103}{{\em Fortsch. Phys.} {\bfseries 70} no.~7-8, (2022) 2200103}, \href{http://arxiv.org/abs/2206.04708}{{\ttfamily arXiv:2206.04708 [hep-th]}}.

\bibitem{Sen:1998sm}
A.~Sen, ``{Tachyon condensation on the brane anti-brane system},'' \href{http://dx.doi.org/10.1088/1126-6708/1998/08/012}{{\em JHEP} {\bfseries 08} (1998) 012}, \href{http://arxiv.org/abs/hep-th/9805170}{{\ttfamily arXiv:hep-th/9805170}}.

\bibitem{Sen:2002nu}
A.~Sen, ``{Rolling tachyon},'' \href{http://dx.doi.org/10.1088/1126-6708/2002/04/048}{{\em JHEP} {\bfseries 04} (2002) 048}, \href{http://arxiv.org/abs/hep-th/0203211}{{\ttfamily arXiv:hep-th/0203211}}.

\bibitem{Sen:2002in}
A.~Sen, ``{Tachyon matter},'' \href{http://dx.doi.org/10.1088/1126-6708/2002/07/065}{{\em JHEP} {\bfseries 07} (2002) 065}, \href{http://arxiv.org/abs/hep-th/0203265}{{\ttfamily arXiv:hep-th/0203265}}.

\bibitem{Kofman:2005yz}
L.~Kofman and P.~Yi, ``{Reheating the universe after string theory inflation},'' \href{http://dx.doi.org/10.1103/PhysRevD.72.106001}{{\em Phys. Rev. D} {\bfseries 72} (2005) 106001}, \href{http://arxiv.org/abs/hep-th/0507257}{{\ttfamily arXiv:hep-th/0507257}}.

\bibitem{Frey:2005jk}
A.~R. Frey, A.~Mazumdar, and R.~C. Myers, ``{Stringy effects during inflation and reheating},'' \href{http://dx.doi.org/10.1103/PhysRevD.73.026003}{{\em Phys. Rev. D} {\bfseries 73} (2006) 026003}, \href{http://arxiv.org/abs/hep-th/0508139}{{\ttfamily arXiv:hep-th/0508139}}.

\bibitem{Frey:2023khe}
A.~R. Frey, R.~Mahanta, A.~Maharana, F.~Muia, F.~Quevedo, and G.~Villa, ``{String thermodynamics in and out of equilibrium: Boltzmann equations and random walks},'' \href{http://dx.doi.org/10.1007/JHEP03(2024)112}{{\em JHEP} {\bfseries 03} (2024) 112}, \href{http://arxiv.org/abs/2310.11494}{{\ttfamily arXiv:2310.11494 [hep-th]}}.

\bibitem{Frey:2024jqy}
A.~R. Frey, R.~Mahanta, A.~Maharana, F.~Quevedo, and G.~Villa, ``{Gravitational Waves from High Temperature Strings},'' \href{http://arxiv.org/abs/2408.13803}{{\ttfamily arXiv:2408.13803 [hep-th]}}.

\bibitem{Aggarwal:2020olq}
N.~Aggarwal {\em et~al.}, ``{Challenges and opportunities of gravitational-wave searches at MHz to GHz frequencies},'' \href{http://dx.doi.org/10.1007/s41114-021-00032-5}{{\em Living Rev. Rel.} {\bfseries 24} no.~1, (2021) 4}, \href{http://arxiv.org/abs/2011.12414}{{\ttfamily arXiv:2011.12414 [gr-qc]}}.

\bibitem{Sarangi:2002yt}
S.~Sarangi and S.~H.~H. Tye, ``{Cosmic string production towards the end of brane inflation},'' \href{http://dx.doi.org/10.1016/S0370-2693(02)01824-5}{{\em Phys. Lett. B} {\bfseries 536} (2002) 185--192}, \href{http://arxiv.org/abs/hep-th/0204074}{{\ttfamily arXiv:hep-th/0204074}}.

\bibitem{Copeland:2003bj}
E.~J. Copeland, R.~C. Myers, and J.~Polchinski, ``{Cosmic F and D strings},'' \href{http://dx.doi.org/10.1088/1126-6708/2004/06/013}{{\em JHEP} {\bfseries 06} (2004) 013}, \href{http://arxiv.org/abs/hep-th/0312067}{{\ttfamily arXiv:hep-th/0312067}}.

\bibitem{Majumdar:2005qc}
M.~Majumdar, ``{A Tutorial on links between cosmic string theory and superstring theory},'' in {\em {COSLAB 2004}}.
\newblock 12, 2005.
\newblock \href{http://arxiv.org/abs/hep-th/0512062}{{\ttfamily arXiv:hep-th/0512062}}.

\bibitem{Conlon:2024uob}
J.~P. Conlon, E.~J. Copeland, E.~Hardy, and N.~S. Gonz\'alez, ``{Percolating Cosmic String Networks from Kination},'' \href{http://arxiv.org/abs/2406.12637}{{\ttfamily arXiv:2406.12637 [hep-ph]}}.

\bibitem{LISACosmologyWorkingGroup:2023njw}
{\bfseries LISA Cosmology Working Group} Collaboration, E.~Bagui {\em et~al.}, ``{Primordial black holes and their gravitational-wave signatures},'' \href{http://arxiv.org/abs/2310.19857}{{\ttfamily arXiv:2310.19857 [astro-ph.CO]}}.

\bibitem{Kallosh:2015nia}
R.~Kallosh, F.~Quevedo, and A.~M. Uranga, ``{String Theory Realizations of the Nilpotent Goldstino},'' \href{http://dx.doi.org/10.1007/JHEP12(2015)039}{{\em JHEP} {\bfseries 12} (2015) 039}, \href{http://arxiv.org/abs/1507.07556}{{\ttfamily arXiv:1507.07556 [hep-th]}}.

\bibitem{Crino:2020qwk}
C.~Crin\`o, F.~Quevedo, and R.~Valandro, ``{On de Sitter String Vacua from Anti-D3-Branes in the Large Volume Scenario},'' \href{http://dx.doi.org/10.1007/JHEP03(2021)258}{{\em JHEP} {\bfseries 03} (2021) 258}, \href{http://arxiv.org/abs/2010.15903}{{\ttfamily arXiv:2010.15903 [hep-th]}}.

\bibitem{AbdusSalam:2022krp}
S.~AbdusSalam, C.~Crin\`o, and P.~Shukla, ``{On K3-fibred LARGE Volume Scenario with de Sitter vacua from anti-D3-branes},'' \href{http://dx.doi.org/10.1007/JHEP03(2023)132}{{\em JHEP} {\bfseries 03} (2023) 132}, \href{http://arxiv.org/abs/2206.12889}{{\ttfamily arXiv:2206.12889 [hep-th]}}.

\bibitem{McAllister:2024lnt}
L.~McAllister, J.~Moritz, R.~Nally, and A.~Schachner, ``{Candidate de Sitter Vacua},'' \href{http://arxiv.org/abs/2406.13751}{{\ttfamily arXiv:2406.13751 [hep-th]}}.

\bibitem{Klebanov:2000hb}
I.~R. Klebanov and M.~J. Strassler, ``{Supergravity and a confining gauge theory: Duality cascades and chi SB resolution of naked singularities},'' \href{http://dx.doi.org/10.1088/1126-6708/2000/08/052}{{\em JHEP} {\bfseries 08} (2000) 052}, \href{http://arxiv.org/abs/hep-th/0007191}{{\ttfamily arXiv:hep-th/0007191}}.

\bibitem{Giddings:2001yu}
S.~B. Giddings, S.~Kachru, and J.~Polchinski, ``{Hierarchies from fluxes in string compactifications},'' \href{http://dx.doi.org/10.1103/PhysRevD.66.106006}{{\em Phys. Rev. D} {\bfseries 66} (2002) 106006}, \href{http://arxiv.org/abs/hep-th/0105097}{{\ttfamily arXiv:hep-th/0105097}}.

\bibitem{Candelas:1989js}
P.~Candelas and X.~C. de~la Ossa, ``{Comments on Conifolds},'' \href{http://dx.doi.org/10.1016/0550-3213(90)90577-Z}{{\em Nucl. Phys. B} {\bfseries 342} (1990) 246--268}.

\bibitem{Baumann:2006th}
D.~Baumann, A.~Dymarsky, I.~R. Klebanov, J.~M. Maldacena, L.~P. McAllister, and A.~Murugan, ``{On D3-brane Potentials in Compactifications with Fluxes and Wrapped D-branes},'' \href{http://dx.doi.org/10.1088/1126-6708/2006/11/031}{{\em JHEP} {\bfseries 11} (2006) 031}, \href{http://arxiv.org/abs/hep-th/0607050}{{\ttfamily arXiv:hep-th/0607050}}.

\end{thebibliography}\endgroup

\end{document}